\documentclass[prb,preprint,amsmath,amssymb,nofootinbib]{revtex4}

\usepackage{color,graphicx}

\def\hkl#1{\hbox{#1}}
\def\dn{\downarrow}
\def\up{\uparrow}

\ifdefined\licInfoON
\def\lic#1#2#3{{{\color{red} [Lic: #1]}}{\color{blue} [{\tt src}: #2\cite{#3}]}}
\else
\def\lic#1#2#3{\relax}
\fi

\begin{document}

\title{Anisotropic magnetoresistance:\\ materials, models and applications}

\author{
Philipp Ritzinger$^{1,2}$ and Karel V\'yborn\'y$^1$}

\address{$^{1}$FZU --- Institute of Physics, Academy of Sciences of the
  Czech Republic, Cukrovarnick\'a 10, Praha 6, CZ--16253\\
$^{2}$MFF --- Faculty of Mathematics and Physics, Charles University, \\
  Ke Karlovu 5, Praha 2, CZ--12000}
\email{ritzinger@fzu.cz,vybornyk@fzu.cz}

\begin{abstract}
Resistance of certain (conductive and otherwise isotropic)
ferromagnets turns  out to exhibit anisotropy with respect to the
direction magnetisation: $R_\parallel$ different from $R_\perp$ with
reference to the electric current direction. This century-old
phenomenon is reviewed both from the perspective of materials and
physical mechanisms involved. More recently, this effect has also been
extended to antiferromagnets. This opens the possibility for
industrial applications reaching far beyond the current ones, e.g.hard
drive read heads. 
\end{abstract}

\maketitle

\section{Intro}
\label{ch_Intro}

Electric resistance $R$ of a conductor depends on the state of its magnetic
order; for example, in ferromagnetic metals at saturation, it depends on the
direction of magnetisation $\vec{M}$. Experimentally, control of external
magnetic field $\vec{B}$ allows to change $\vec{M}$ and this suggests the
name {\em magnetoresistance}. The reader should not be misled into
thinking that any dependence $R(B)$ is confined to magnetically
ordered materials though. Magnetoresistances encompass a wide range of
phenomena and in this review, we only focus on situations where the
{\em anisotropy} of $R$ is caused by magnetic order. By large part, we
will discuss ferromagnets (FMs) where such anisotropic magnetoresistance (AMR) 
has been explored extensively but only few reviews exist and the most popular
Potter and McGuire~\cite{McG-Potter} article is now almost half a century old.
More modern developments of the field will also be discussed where as there
seems to be a shift of focus from FMs to materials with more complex
magnetic order (of which antiferromagnets are of particular interest) and
here, even an elementary consensus on terminology is still to be reached.

After this introductory Section, we turn our attention to approaches
to model and thus understand the AMR on phenomenological and microscopic level
(Sec. 2) and then, to materials where
AMR has been explored (Sec.~3). AMR applications are discussed in
Sec.~4 and the last section is devoted to conclusions.

\subsection{Basic observations}
\label{sec_Def}

The basic approach to quantify AMR in a given ferromagnetic material
is to compare resistance for magnetisation parallel and perpendicular to
current direction relative~\cite{Isnaini:2020_a} to their suitably
chosen average $R_0$:
\begin{equation}\label{eq-01}
  \mbox{AMR}=\frac{R_\parallel-R_\perp}{R_0}.
\end{equation}
Depending on context, the most obvious choice $R_0=(R_\parallel+R_\perp)/2$
may be replaced by another weighted sum~\cite{McG-Potter} but since AMR
is typically of the order of per cent, this is usually of little consequence.
AMR in most metals is positive and it depends on temperature: it vanishes
when magnetic order is lost upon heating.  

A more careful analysis of AMR requires the consideration of full
resistivity tensor $\rho_{ij}$. In a single crystal (of sufficiently
low\footnote{Note that even if all $\rho_{ii}$ components are
equal in a cubic crystal when $\vec{M}=0$, additional crystalline AMR
terms appear (unlike for polycrystals) once magnetisation is
taken into account.} symmetry), anisotropies appear already for $\vec{M}=0$
and the AMR must not be confused with these 'fundamental anisotropies'.
Even cubic systems can, however, exhibit non-zero off-diagonal components
of $\rho_{ij}$ under non-zero magnetisation (see Sec.~1c) and to this end,
angular dependence of $\rho$ should be considered; the most common
observation is
\begin{equation}\label{eq-02}
  \rho_{xx}/\rho_0 = 1+C_I\cos 2\varphi
\end{equation}
where $C_I$ is sometimes called non-crystalline AMR because it survives
(as opposed to the more complex angular dependences discussed in
Sec.~\ref{ch_Model}\ref{sec_modelpheno} for example) even in polycrystalline
systems. Eq.~\ref{eq-02} can be obtained by averaging expressions
such as Eq.~\ref{F_Prague} over orientations of crystallites whereupon
only $\varphi$ (angle between current and $\vec{M}$) remains invariant.
Clearly the AMR as defined in Eq.~\ref{eq-01} is just twice $C_I$
when no crystalline AMR is present.

\begin{figure}[!h]
\centering\includegraphics[width=5in]{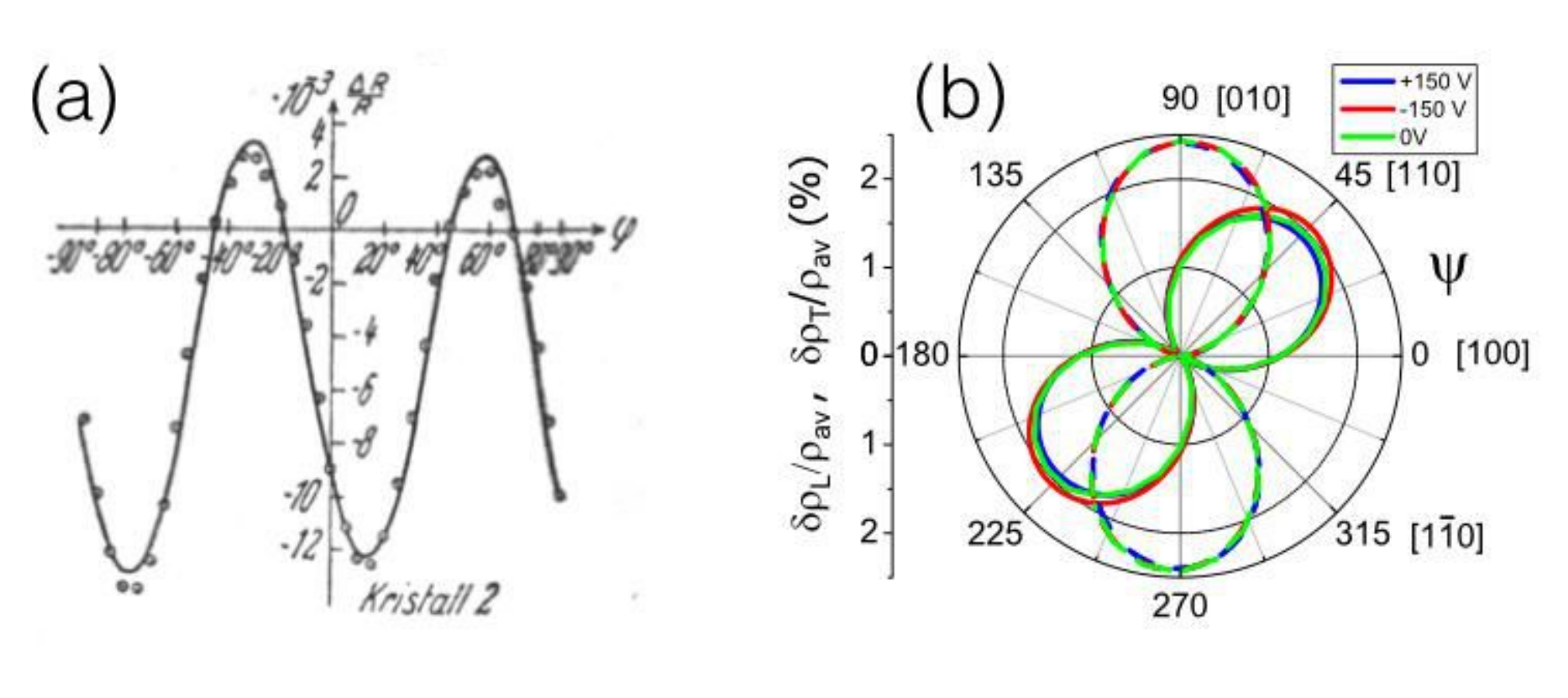}
\caption{\lic{YES}{D\"oring+Elisa}{f01l}
  Two basic examples of AMR measurement. (a) D\"oring's measurements
  on nickel crystals~[\onlinecite{Doring:1938}], (b) longitudinal and
  transversal AMR measured on 
  (Ga,Mn)As thin films~[\onlinecite{deRanieri:2008_a}]. Reproduced
  with permission of John Wiley and Sons.}
  \label{fig-01}
\end{figure}

\subsection{History and More Features}
\label{sec_History}

\textit{Discovery.} Transition metals were the first materials where AMR was
discovered: in 1857, William Thomson measured~\cite{Thomson} in Fe and Ni what
we would call non-crystalline AMR nowadays. The discovery in the third elemental
room temperature ferromagnet, Co, was made a little later~\cite{Thomlinson:1882}
While these measurements concerned polycrystalline samples, D\"oring
in 1938 investigated~\cite{Doring:1938}  the AMR in Fe and Ni single-crystals
more thoroughly as a function of $\varphi$ and also the angle of $\vec{M}$
respective to crystallographic directions. Apart from the non-crystalline
AMR~(\ref{eq-02}) terms dependent on crystal symmetry (crystalline
AMR) were found. His phenomenological approach to describe the full
AMR in single crystals is still frequently used in modern
works~\cite{Limmer:2006, Li:2010, deRanieri:2008_a, Althammer, Ritzinger:2021}
as discussed in sec. \ref{ch_Model}\ref{sec_modelpheno}.\\

\textit{Intrinsic and Extrinsic AMR.}
Next to the possible classification into non-crystalline and
crystalline AMR, we can also make the distinction between intrinsic
and extrinsic contributions. In the simplest case of Drude formula,
\begin{equation}
  \sigma_0=\frac{ne^2\tau}m=\omega^2_p\epsilon\tau,\qquad
  \sigma(\omega)=\frac{\sigma_0}{1-i\omega\tau}
  \label{eq-05}
\end{equation}
the extrinsic (thus scattering-dependent) effects enter through the
the dependence of relaxation time $\tau$ on the magnetisation direction
while the intrisic contribution to AMR amounts to such a dependence of
the plasma frequency $\omega_p$. Examples of the former mechanism can be
captured by effective models described in Sec. 2(b)i and the prime example
is s-d-scattering, thus, that delocalized conduction electrons (4s)
are scattering into localized 3d states via magnetic impurities. 
Intrinsic AMR receives more attention in recent years~\cite{Zeng:2020_a},
since investigated materials are generally more complicated and
bandstructure-calculation has become more precise, allowing for a more
thorough distinction. On a theoretical side, AMR can be calculated
from the bandstructure (intrinsic contribution) and is then compared
to experimental results. If there happens to be a significant
difference, this can be attributed to scattering (extrinsic
contribution). Experimentally, the usage of AC-voltage can be used
to distinguish~\cite{Nadvornik:2021_a} the intrinsic and extrinsic
contributions to $\sigma(\omega)$ in Eq.~(\ref{eq-05}), since the
intrinsic contribution leads to frequency-independent term in AMR,
while the extrinsic contribution scales with $\propto 1/\omega$, see
Sec. 4(b)i for details. \\

\textit{Negative AMR.} In most common metals, AMR as defined by
Eq.~(\ref{eq-01}) is positive; this is fairly demonstrated by
Tab.~\ref{table-TM} which shows also one of early examples of systems
where AMR is negative (cobalt with traces of iridium). The first
materials where negative AMR was found were, nevertheless, much more
common alloys of transition metals with aluminium~\cite{Bates:1946_a}.
The  belief that negative AMR is an exception established itself in the next couple of decades, which may be fueled by the fact that major theories of AMR were developed on simple transition metals showing positive AMR under normal circumstances.




\begin{figure}[b]
  \begin{center}
    \begin{tabular}{c|c|c}
      &$\rho_\dn>\rho_\up$  & $\rho_\up>\rho_\dn$ \\ \hline
    $d_\up>d_\dn$ & (c) neg. & (a) pos. \\ \hline
    $d_\up<d_\dn$ & (b) pos. & (d) neg.
  \end{tabular}
  \end{center}
	\caption{\lic{NN}{Kokado}{f02l}   
    Sign of AMR explained in the context of $sd$ model. Examples:
    (a) bcc Fe, (b) fcc Co or Ni, (c) half-metallic ferromagnets such as Co$_2$MnAl$_{1-x}$Si$_x$, and (d) Fe$_4$N. Inspired
    heavily by Fig.~4 of Kokado et al.~\cite{Kokado:2012}.
  }
  \label{fig-02}
\end{figure}

One of the main approaches to microscopically understand the AMR, so 
called $sd$-model which is explained in Sec.~2(b)i, allows to understand
the AMR sign (in some materials) using the following simplified picture
based on Mott's two-current model~\cite{Mott-2CM}
which operates with two spin channels and their resistivities
$\rho_{\up,\dn}$. We will follow explanations by Kokado, Tsunoda et
al.~\cite{Kokado:2012}, where the DOS at Fermi level $E_F$ in the
majority/minority $d$-bands is $d_\up/d_\dn$. As Fig.~\ref{fig-02} shows,
majority spin conduction (hence $d_{\up} > d_{\dn}$ and $\rho_{\up} > \rho_{\dn}$ and vice versa) is responsible positive AMR, while minority spin conduction ($d_{\up} < d_{\dn}$ and $\rho_{\up} > \rho_{\dn}$ and vice versa) causes negative AMR. 
The key parameter is thus $\alpha=\rho_\dn/\rho_\up$ and a detailed 
discussion~\cite{Kokado:2012} serves as a useful guideline for the AMR
sign across the whole material class of transition metals. Validity of
this guideline is limited, however, by the range of applicability of
the $sd$-model: other material classes, such as dilute magnetic
semiconductors discussed in Sec.3(b), follow different
patterns~\cite{Rushforth:2007_a,Vyborny:2009_a}.


In the context of this theory~\cite{Kokado:2012}, negative AMR is sometimes
promoted to be a possible sign of half-metallicity \cite{Yang:2012, Sato:2018, Sato:2019},
which has to be taken with caution: first, the sign of AMR as defined by
Eq.~(\ref{eq-01}) may depend on the current direction with respect to
the crystal (in which case it makes better sense to analyse AMR in
terms of its non-crystalline and crystalline components, {\em see Sec.~2a})
and this clearly cannot mean that the system would be half-metal in one
case and normal metal in the other case. An example of a material
which is clearly {\em not} a half-metal is the 30:70 alloy of iron and
cobalt~\cite{Miao:2021} (sign change of AMR can be seen in Fig.~2 of that
reference where $\vec{B}$ and $\vec{M}$ are nearly parallel). Also, temperature
variation can cause similar changes (e.g. in Mn$_4$N \cite{Kabara:2017}). Second
even in predominantly negative signed Co-based Heusler alloys, positive AMR was reported by e.g. variations of the stoichiometry \cite{Sato:2019} or the annealing temperature \cite{Yang:2013} (see sec. \ref{sec_Heusler}). The changes of sign in all of these materials were explained successfully within the framework of the aforementioned majority/minority scattering by Kokado and Tsunoda.

Still, it holds true that half-metallic density of states induces
negative sign of AMR. The backwards conclusion (negative sign implies
half-metallicity \cite{Miao:2021}) is not generally true. Other systems
where AMR can be negative will be discussed later throughout this review: 
certain antiferromagnets, manganites, two-dimensional electron gases
to name a few.

\subsection{AMR and the more fancy effects}
\label{sec_similareffects}

\begin{figure}[!h]
\includegraphics[scale=0.25,angle=0]{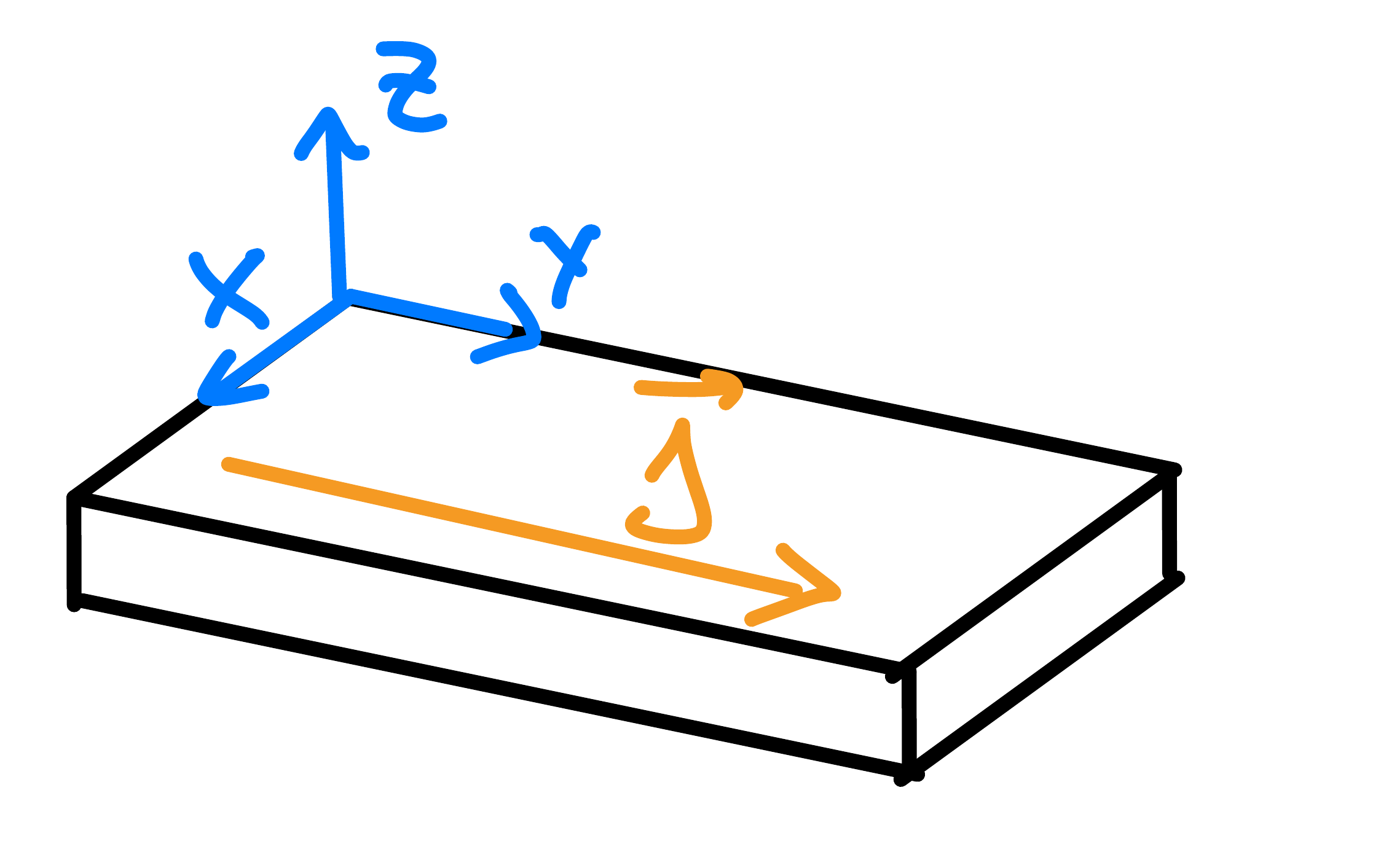}
\includegraphics[scale=0.45,angle=0]{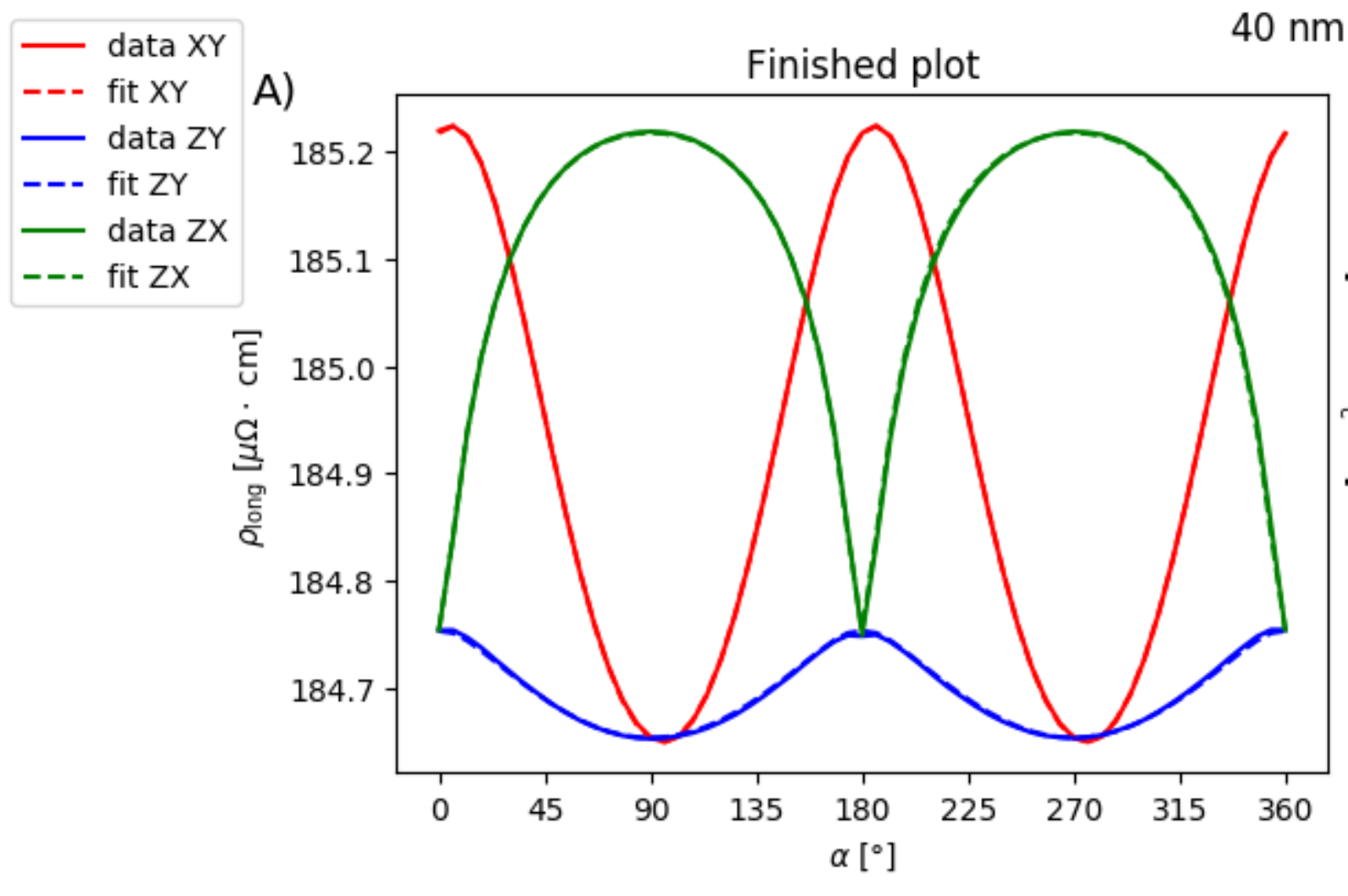}
\caption{\lic{NN}{Philipp}{f03l}
  Example of Stoner-Wohlfahrth analysis in AMR data of a Co$_2$MnGa thin-film sample. Alongside the SW1 model and a basic non-crystalline AMR, also higher-order crystalline AMR terms are taken into account (see sec.~\ref{sec_modelpheno}). There is an excellent agreement between data and fit. The magnetic field was rotated in three different rotation planes denoted as XY, ZY and ZX, where Z$ = \hat{n} \parallel [001]$, Y$ = \hat{j} \parallel [110]$ and X = $Y \times Z$. The rotation in the XY plane begins at the X axis and in the other plane at the Z axis. Reproduced from Ref.~\cite{Ritzinger:2020}.}
    \label{fig-03}
\end{figure}

We first wish to elucidate the relationship of AMR to off-diagonal
component of the resistivity tensor,
\begin{equation}
  \rho_{xy}/\rho_0=C_I\sin 2\varphi
  \label{eq-09}
\end{equation}
%
in the simplest case, which is often called the planar Hall effect.
Assume a planar system with magnetization $\vec{m}\parallel x$ which would
be otherwise isotropic (in other words, $\vec{m}$ provides the only
source of symmetry breaking). Let us denote the two non-zero
components $\rho_{xx}$ and $\rho_{yy}$ by $\rho_\parallel$ and
$\rho_\perp$, respectively. Now consider a rotation of $\vec{m}$ to 
$R_\phi\vec{m}$: in a polycrystal, this would be equivalent to leaving
$\vec{m}$ unchanged and rotating the resistivity tensor instead:
\begin{equation}
  R_\phi\left(\begin{array}{cc}\rho_\parallel & 0 \\ 0 & \rho_\perp\end{array}
    \right)R^T_\phi
   =
   \left(\begin{array}{cc}\rho_0+\frac12 \Delta\rho \cos 2\phi & \frac12 \Delta\rho \sin 2\phi \\
     \frac12 \Delta\rho \sin 2\phi & \rho_0-\frac12 \Delta\rho \cos 2\phi \end{array}
    \right)
\label{eq_rotrho}
\end{equation}
where $\Delta\rho=\rho_\parallel-\rho_\perp$ and $R_\phi$ is an
orthogonal matrix. The off-diagonal
elements can be identified with Eq.~(\ref{eq-09}) and it is therefore
appropriate to call that effect (i.e. PHE) the transversal AMR. As a remark
we point out that 'transverse AMR' is sometimes used~\cite{Kabara:2016}
to describe the experimental configuration where magnetisation rotates
in the plane perpendicular to the current direction ({\em green curve}
shown in Fig.~\ref{fig-03}); in Eqs.~(\ref{eq-02},
\ref{eq-09}) this corresponds to constant $\phi=\pi/2$ and one would then
naively expect no variation of resistance. We explain in Sec.~2(a) that
{\em crystalline AMR} is responsible for any signal measured in this setup.

AMR belongs to a wider family of transport phenomena in magnetically ordered
materials and in the following we mention several further examples of
its members. They are all bound by Onsager reciprocity relations, for
resistitivy tensor they read
%
\begin{equation}
  \rho_{ij}(M,B)=\rho_{ji}(-M,-B)
\end{equation}
and to begin with, we observe that for $\rho_{xy}$, this relation can
be fulfilled either by Eq.~(\ref{eq-09}) in the transverse AMR
(a symmetric tensor component $\rho_{xy}=\rho_{yx}$ which is even in
magnetisation) or by the anomalous Hall effect (AHE) with $\rho_{xy}=-\rho_{yx}$
odd in magnetisation. Next, there are thermoelectric counterparts
of these effects, the anomalous Nernst effect (to AHE) and the anisotropic
magnetothermopower discussed in Sec. 4(b)iii. Spin conductivity
instead of charge conductivity can also be studied (e.g. SHE instead
of AHE) and both effects are closely related~\cite{Omori:2019_a}, e.g.
in permalloy AHE scales with the spin Hall effect (SHE) in proportion
to the spin polarisation.
Finally, we wish to mention transport in ballistic rather
than diffusive regime: tunneling AMR (TAMR) and ballistic AMR
discussed in Sec. 4(b)iv.

\subsection{What is AMR and what it is not}
\label{sec_notAMR}

Magnetoresistance (MR) may refer to any phenomenon~\cite{Pippard:1989} 
where $R(B)$ is not constant and as such they are not limited in scope to
magnetically ordered materials. Orbital effects leading to MR imprint
the anisotropy of crystal to $R(B)$, as recently nicely reviewed by
Zhang et al.~\cite{Zhang:2019_a}, and ensuing anisotropic MR is {\em not}
the subject of the present review.

On the other hand, the AMR appears under different names in
literature: spontaneous magnetoresistance anisotropy (SMA)~\cite{Ebert:1996_a},
spontaneous resistivity anisotropy (SRA)~\cite{Stampe:1995_a,Ziese:2000_a},
ferromagnetic anisotropy of resistivity (FAR)~\cite{Chakraborty:1998_a}
or magneto-resistivity anisotropy~\cite{Alagoz:2014}.
Also, longitudinal MR and transversal MR
are sometimes discussed separately~\cite{El-Tahawy:2022} whereas their
difference in high magnetic field is the actual AMR.
In some occasions, the term AMR or anisotropic MR is used, when the MR ratio is plotted for different field directions \cite{Mizusaki:2009_a, Bolte:2005}. In that case it can be, that the AMR ratio is not quantitatively calculated as in Eq.~\ref{eq-01}, but the discussion is rather restricted to the mere fact, that the MR is different for different field direction, thus implying AMR.
Ideally, we are interested in magnetically ordered materials at saturation.\\

\textit{Misconception with MCA.} A frequent trouble is the confusion of AMR and magneto-crystalline Anisotropy (MCA). Whenever there is a deviation from the classical two-fold dependence $\Delta \rho \propto \cos^2(\sphericalangle (\textbf{H, J}))$ (where \textbf{H} and \textbf{J} is the current density) it is not per se clear whether they stem from MCA or are AMR terms. MCA can lead to higher-order symmetries on the AMR signal, however these terms might also origin from AMR due to crystalline symmetry (so called \textit{crystalline AMR} or \textit{single-crystal AMR} (SCAMR)). The frequent conclusion, the higher-order terms are steming from MCA is only unequivocally true in polycrystalline materials. In single-crystals a careful distinction of these MCA and AMR is always mandatory (e.g. by determining the value of MCA in a different experiment and account for it). Furthermore it should be kept in mind that crystalline AMR and MCA are not the same effect: While both are dependent on the bandstructure, a key ingredient of any (extrinsic) AMR is scattering, which does not play a role in MCA. The intrinsic AMR depends on the anisotropy of the fermi velocities, which is not necessarily linked to the exchange energy causing the MCA into existence. The concept of MCA is further elaborated in section \ref{sec_stepone} and the crystalline AMR in derived and explained in detail in section \ref{sec_TM}. An illustration of the difference of AMR and MCA can be seen in Fig. 4d-g of \cite{Alagoz:2014}, where the AMR and MCA show much different temperature dependences.

\subsection{Step one: magnetisation control}
\label{sec_stepone}

As already explained, it is usually the applied magnetic field $\vec{B}$
that steers the magnetic moments. In other words, it is desirable to
determine the magnetic state depending on $\vec{B}$. First, we assume that
we are looking at a single-domain state (effects related
to a non-trivial domain structure tend to be more severe in
antiferromagnets~\cite{Kriegner:2016_a}); next,
we want to focus only on classical
magnetism. Under these assumptions we are basically left
with inter-sublattice exchange coupling (if there's more than just one
magnetic sublattice) and magnetic anisotropy.

A convenient framework in ferromagnets is the time-proven
Stoner-Wohlfarth model~\cite{Stoner:1948_a} (henceforth referred
to as the SW1 model), summarised by Eq.~2 in~\cite{Volny:2020_a},
which yields the local energy minimum for magnetisation depending on
history and two parameters: $B=|\vec{B}|$ and $B_a$ (magnetic anisotropy).
Among others, SW1 models are widely used in the analysis of
resistivity data. An example can be found in Fig.~\ref{fig-03}, where
next to basic non-crystalline AMR, a SW1 model and higher order
crystalline AMR components were taken into account (see Sec
2(a) for the latter); the latter become manifest in (a) the different 
amplitudes of the blue and red curves or (b) non-constant signal plotted
as the green curve. As for (b), magnetisation remains always perpendicular
to current, $\varphi=\pi/2$, and if Eq.~(\ref{eq-02}) were the complete
description of AMR in this case, $\rho_{yy}$ should remain
constant. By including {\em crystalline AMR} terms into Eq.~(\ref{eq-02}) as
discussed later, see Eq.~(\ref{F_Prague}), the observed behaviour both
for (b) and (a) can be well understood.
The same type of description (based on SW1, see Fig.~\ref{fig-10}) was used
by Limmer et al \cite{Limmer:2006, Limmer:2008} for (Ga,Mn)As.

As soon as there are more than one magnetic sublattice (MSL), the situation
becomes less straightforward~\cite{Song:2018_a}. It is possible to
generalise the previous approach to antiferromagnets with two MSLs: such
SW2 model reads
\begin{equation}\label{eq-03}
  \frac{E}{MV}= B_e\vec{m}_1\cdot\vec{m}_2 -B\vec{b}\cdot(\vec{m}_1+\vec{m}_2)
  +B_a [(\vec{m}_1\cdot \hat{a})^2+(\vec{m}_2\cdot \hat{a})^2].
\end{equation}
and a new parameter has been introduced: the inter-sublattice exchange coupling
$B_e$. The basic mode of operation of SW2~\cite{Correa:2018_a} is that the
N\'eel vector $\vec{L}=\vec{m}_1-\vec{m}_2$ is perpendicular to $\vec{B}$ 
which always (for $|B|>0$) corresponds to energy minimum in
Eq.~(\ref{eq-03}) once $B_a=0$. In this way, $\vec{L}$ can be
effectively controlled by $\vec{B}$ and for finite $B_a$, the same applies
beyond spin-flop field $\propto\sqrt{B_aB_e}$.

This concept can be extended to more complicated systems and starting with
SW3, non-collinear magnetic order has to be considered. Recently, Mn$_3$X
materials (where X can be Ge or Sn, for example) attracted significant
attention and Liu and Balents~\cite{Liu:2017_a} discuss a model where
beyond adding a third MSL to Eq.~(\ref{eq-03}) also Dzyaloshinskii-Moriya
interaction is included. Geometry of kagome lattice (see Fig.~\ref{fig-16})
introduces frustration and relationships between $\vec{B}$ and $\vec{m}_{1,2,3}$
are in general difficult to describe in simple terms.


\section{Modelling}
\label{ch_Model}

In this chapter, the different modelling approaches are presented. We
will start in sec. \ref{sec_modelpheno} by introducing potent
phenomenlogical models, which allow to effectively analyse the even most
complex AMR data. Due to their phenomenological nature, however, they
cannot give insight about the possible origins of individual terms in
expansions such as~(\ref{F_Prague}). While more involved, microscopical 
models reviewed in Sec.~\ref{sec_modelmicro} make such deeper understanding
possible.

\subsection{Phenomenlogical models}
\label{sec_modelpheno}
We define the magnetic field direction to be $\textbf{h} = \textbf{H} / H$ and the magnetization direction to be $\textbf{m} = \textbf{M} / M$. Please keep in mind that the AMR depends on $\textbf{m}$ and not on \textbf{h} - the rotation of the magnetic field is simply used to control the rotation of the magnetization. The dependence of \textbf{m} on \textbf{h} was discussed in the previous section and the confusion of MCA with AMR in sec. \ref{sec_notAMR}. Broadly speaking it holds that: $AMR = \rho(\textbf{m}) \neq \rho(\textbf{h})$.\\
The simplest possible way to describe the AMR presents itself as~(\ref{eq-02}):
$\Delta\rho(\textbf{m}) \propto \cos(2 \varphi)$, where $\varphi$ is the angle between \textbf{m} and current direction $\textbf{j} = \textbf{J}/J$. In a single-crystal this simple picture does not hold anymore, but instead the AMR can have more complex contributions depending on the crystalline symmetry. In the following section, we will present a simple yet extremely powerful phenomenological model to describe (however \textit{not} explain) even complex AMR data, which was originally developed by D\"oring in 1938 \cite{Doring:1938} and since then used many times again \cite{Limmer:2006, Limmer:2008, deRanieri:2008_a, Althammer, Li:2010, Ritzinger:2021}. \\ 

\textit{The model.} To begin with, we assume that we do not know the correct analytical expression of the resistivity $\rho$ and that $\rho$ depends only on the direction of the magnetization \textbf{m}. Furthermore, there can be higher-order dependencies on \textbf{m}. Thus, we express $\rho$ as a power series of \textbf{m}:

\begin{equation}
\rho_{\mathrm{ij}}(\hat{m}) = \rho_{\mathrm{ij}}^{(0)} +  \rho_{\mathrm{ijk}}^{(1)} m_{\mathrm{k}} +  \rho_\mathrm{{ijkl}}^{(2)} m_{\mathrm{k}} m_\mathrm{{l}} +  \rho_{\mathrm{ijklm}}^{(3)} m_{\mathrm{k}} m_{\mathrm{l}} m_{\mathrm{m}} + \rho_{\mathrm{ijklmn}}^{(4)} m_{\mathrm{k}} m_{\mathrm{l}} m_{\mathrm{m}} m_{\mathrm{n}} + ... 
\label{eq_expansion}		
\end{equation}
 
 where $\rho_{\mathrm{ij}}^{(0)}$, $\rho_{\mathrm{ijk}}^{(1)}$, $\rho_{\mathrm{ijkl}}^{(2)}$, $\rho_{\mathrm{ijklm}}^{(3)}$ and $\rho_{\mathrm{ijklmn}}^{(4)}$ are the expansion coefficients and
 	 $m_{\mathrm{k}}, m_{\mathrm{l}}, m_{\mathrm{m}}, m_{\mathrm{n}} \in \{m_{\hkl[1 0 0]}, m_{\hkl[0 1 0]}, m_{\hkl[0 0 1]} \}$ are the cartesian components of \textbf{m}. 
 
 The number of independent parameters is reduced by using the following four strategies: (i) Commutation $m_{\mathrm{k}} m_{\mathrm{l}} = m_{\mathrm{l}} m_{\mathrm{k}}$ for all $m_{\mathrm{k}}$ and $m_{\mathrm{l}}$, (ii) the identity $\textbf{m}^2 = \sum_k m_{\mathrm{k}}^2 = 1$, (iii) the Onsager relation \cite{Althammer}: $\rho_{\mathrm{ij}}(\hat{m}) = \rho_{\mathrm{ji}}(-\hat{m})$ and (iv) Neumann's principle: The resistivity tensor, as well as its expansion coefficients, must reflect the crystal symmetry \cite{Birss:1966}. There are several ways to account for the symmetry, e.g. by using generator matrices of the crystal symmetries is shown in \cite{Limmer:2006, Limmer:2008, Althammer}.

 \begin{figure}[!h]
   \centering\includegraphics[scale=0.45,angle=0]{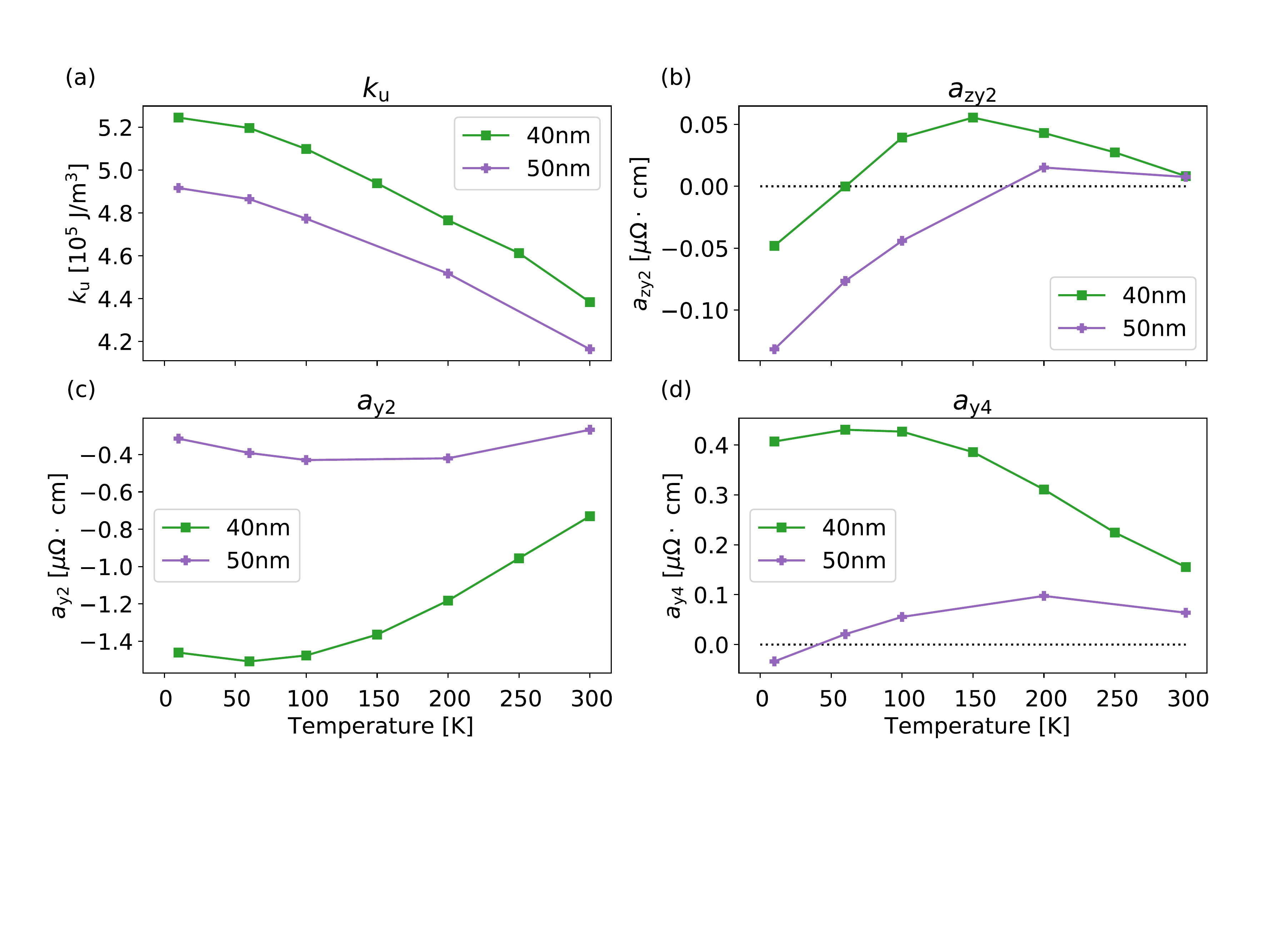}
\caption{\lic{YES}{Philipp}{f09l}
  Temperature evolution of the phenomenological parameters obtained by
  the fit to AMR data of two Co$_2$MnGa thin-film samples. (a)
  Uniaxial magnetic anisotropy from the SW1 model and (b???d) part of
  the parameters for AMR in tetragonal symmetry similar to those
defined in Eq. \ref{eq_long010C1}.
Reproduced from Ref.~\cite{Ritzinger:2021}.}
    \label{fig-09}
\end{figure}

For a more detailed treatment it can be wise to consider the
previously mentioned publications, especially the treatment in
\cite{Althammer}. Next, we explain how the great number of
coefficients appearing in Eq.~(\ref{eq_expansion}) can be reduced to a
small set of key parameters such as those shown in Fig.~\ref{fig-09} for
a specific tetragonal system.

Please note, that this approach yields an expression for the resistivity tensor $\rho_{ij}$ differing depending on the crystal symmetry. The tensor in cubic symmetry can (among others) be found in Eq. 4 of \cite{Limmer:2008} and in tetragonal symmetry in Eq. 4 and 5 of \cite{Limmer:2008}. The resulting tensor depends generally on the components $m_k$ of \textbf{m} and also on coefficients $A, B, ...$ which are unknown in the general case and are sample-dependent.

The longitudinal resistivity $\rho$ is obtained by applying Ohm's law: $\rho = \textbf{j} \rho_{ij} \textbf{j}$. The coefficients of the resistivity do change depending on the crystal symmetry and on the current direction. As an example, $\rho$ in cubic symmetry with $\textbf{j} \parallel \hkl[100] \equiv j_x$ writes as:

\begin{align}
\rho &= \rho_{0} + a_{\mathrm{x2}} \cdot m_x^2  
+ a_{\mathrm{x4}} \cdot m_x^4  
+ a_{\mathrm{zy2}} \cdot m_z^2 \cdot m_y^2	 \\
& = \rho_0 + a_{\mathrm{x2}} \cos(\phi)^2 \sin(\theta)^2  
+ a_{\mathrm{x4}} \cos(\phi)^4 \sin(\theta)^4  
+ a_{\mathrm{zy2}} \sin(\phi)^2 \sin(\theta)^2 \cos(\theta)^2
\label{eq_long010C1}
\end{align}
where the $a_{x2}, a_{x4}, a_{zy2}$ are effective sample-dependent
coefficients, which are linked to the original set of coefficients $A,
B, ...$ and in the second step a parametrization of \textbf{m} in
polar coordinates $\textbf{m} = (\cos(\phi) \sin(\theta), \sin(\phi)
\sin(\theta), \cos(\theta))$ was applied. The calculations are lengthy
and can be found elsewhere, alongside with expressions for the
longitudinal resistivity for current along \hkl[110] or the
resistivity tensor for tetragonal crystal symmetry
\cite{Ritzinger:2021, Limmer:2006, Limmer:2008,
  Althammer}. Expressions for other symmetries in literature are not
known to us. Please note, that the same approach is valid in order to
investigate transversal resistivity, thus Hall effect. An example for
the temperature evolution of some of these phenomenological
coefficients in two Co$_2$MnGa thin-film samples with tetragonal
symmetry can be found in Fig.~\ref{fig-09}.

Eq. \ref{eq_long010C1} is only one possible way of writing things down. For example, D\"oring \cite{Doring:1938} expresses the resistivity in terms of direction cosines of the magnetization $\alpha_i$ and of the current $\beta_i$. Another way of describing the AMR is given by: \cite{deRanieri:2008_a}

\begin{equation}
\frac{\Delta \rho_{\mathrm{long}}}{\rho_\mathrm{av}} = \underbrace{C_{\mathrm{I}} \cdot \cos({2 \varphi})}_\text{non-crystalline} + \overbrace{C_{\mathrm{U}} \cdot \cos({2 \psi})}^\text{uniaxial crystalline} + \underbrace{C_{\mathrm{C}} \cdot \cos({4 \psi})}_\text{cubic crystalline}  + \overbrace{C_{\mathrm{IC}} \cdot \cos({4 \psi - 2 \varphi})}^\text{mixed non-crystalline/crystalline}	
\label{F_Prague}
\end{equation}
where $\varphi$ is the angle between \textbf{m} and \textbf{j} and $\psi$ is the angle between \textbf{m} and a certain, fixed crystallographic direction in the plane of rotation. Eq. \ref{F_Prague} is consistent with the previously shown ansatz Eq. \ref{eq_long010C1}, which was shown in \cite{Ritzinger:2020}. However Eq. \ref{F_Prague} is only a two-dimensional equation (\textbf{m} rotated in the plane of the surface), while Eq. \ref{eq_long010C1} is a three-dimensional equation (AMR can be described for arbitary \textbf{m} on a spherical surface via $\phi$ and $\theta$. In all cases saturation is implied, thus the length of \textbf{M} is irrelevant.) The usuage of two angles $\phi$ and $\psi$ is slightly confusion since it implies three-dimensionality, however $\phi$ and $\psi$ are not defined with respect to different spatial dimensions but instead to different reference axis in order to distinguish between the so-called crystalline and non-crystalline case.\\

\textit{Higher-order contributions}
 are due to crystal structure and thus only appearing in single-crystals or epitaxial materials with sufficient crystal quality. In polycrystalline materials, the AMR will be two-fold (see Eq. \ref{eq-02}) as can be shown by theoretically by averaging the resistivity tensor over all possible crystal orientations (see \cite{Althammer, Limmer:2006, Limmer:2008}) - or even simpler, to set $\psi \equiv 0$ in Eq. \ref{F_Prague} since crystalline directions do not have any meaning in the polycrystalline limit. In doing so one will recover Eq. \ref{eq-02}. This emphasises the usage of the terms non-crystalline (= independent of crystal structure and thus two-fold) and crystalline AMR. \\
 The origin of the crystalline AMR is still under active investigation. While many studies restrict themselves to the mere existence of e.g. a four-fold symmetry, the picture is more complex since the AMR consists of many contributions in various crystalline direction as can be seen above (and e.g. in \cite{Doring:1938, Limmer:2006, Limmer:2008, Li:2010, Althammer, Ritzinger:2021}. While these studies are an accurate description of all the terms possibly existing in the AMR, microscopic studies are rare. For the case of a four-fold symmetry, the effective model of Kokado and Tsunoda \cite{Kokado:2015} (and see following section) showed that a tetragonal symmetry is needed for the four-fold term to appear. The appearance of four-fold terms in many technically cubic materials can be linked to tetragonal distortions induced to thin-films by many substrates.\\
 However, a study describing all the terms in Eq. \ref{eq_long010C1} as well a study for even higher order terms, is still missing to date.\\
While relatively rare, higher order crystalline terms have also been
reported. In hexagonal crystal structures, six-fold AMR can emerge. This
was reported for instance in antiferromagnetic MnTe~\cite{Kriegner:2017_a},
but also in two-dimensional electron gases on hexagonal [111] interfaces 
between transition-metal oxides as discussed further in Sec.~3(e).
 The highest symmetry reported is a eight-fold symmetry measured in (Ga,Mn)As \cite{DeRanieri:2008_a} and in (In,Fe)As \cite{PhanNamHai:2012_a}. In the latter case it was explained by crystal field effects due to a zinc-blende structure. \\
 

\subsection{Microscopic models}
\label{sec_modelmicro}

Regardless of the detailed structure of a microscopic model aiming to describe
AMR in a particular material, two basic ingredients are needed: reasonably
accurate knowledge of the electronic structure and that of momentum relaxation.
On the level of Eq.~\ref{eq-05}, this was reduced to the plasma frequency
which can be evaluated, see Sec. 2b(i), from electron dispersion $E_{\vec{k}}$
%
\begin{equation}
  \omega_p^2=8\pi^2 \hbar^2e^2 \int \frac{d^3 k}{(2\pi)^2} 
  v_x^2 \delta(E_{\vec{k}}-E_F)
  \label{eq-06}
\end{equation}
and, regarding the momentum relaxation, to transport relaxation time 
which can be accessed through the Fermi golden rule
\begin{equation}
\frac1{\tau}=\frac{2\pi n_{imp}}\hbar \int dk'  \delta(E_i-E_f) |M_{kk'}|^2
  (1-\cos \theta_{kk'})
  \label{eq-07}
\end{equation}
whereas we only consider scattering on static disorder (such as point
defects in crystal with density $n_{imp}$).
In the following, we elaborate on two possible strategies to treat
both these ingredients and even if Eqs.~(\ref{eq-06},\ref{eq-07})
represent only {\em examples} of how electronic structure and scattering
can be taken into account, any microscopic model of AMR must in some
way consider them both. We proceed to explain effective models whereas
symbols appearing in the preceding equations will also be described.
Our focus will be, in general, on systems with metallic conduction and
other situations (such as hopping conduction or systems with bound
magnetic polarons~\cite{Terry:1996_a}) will not be discussed in this review.


\begin{figure}[!h]
  \includegraphics[scale=0.3,angle=0]{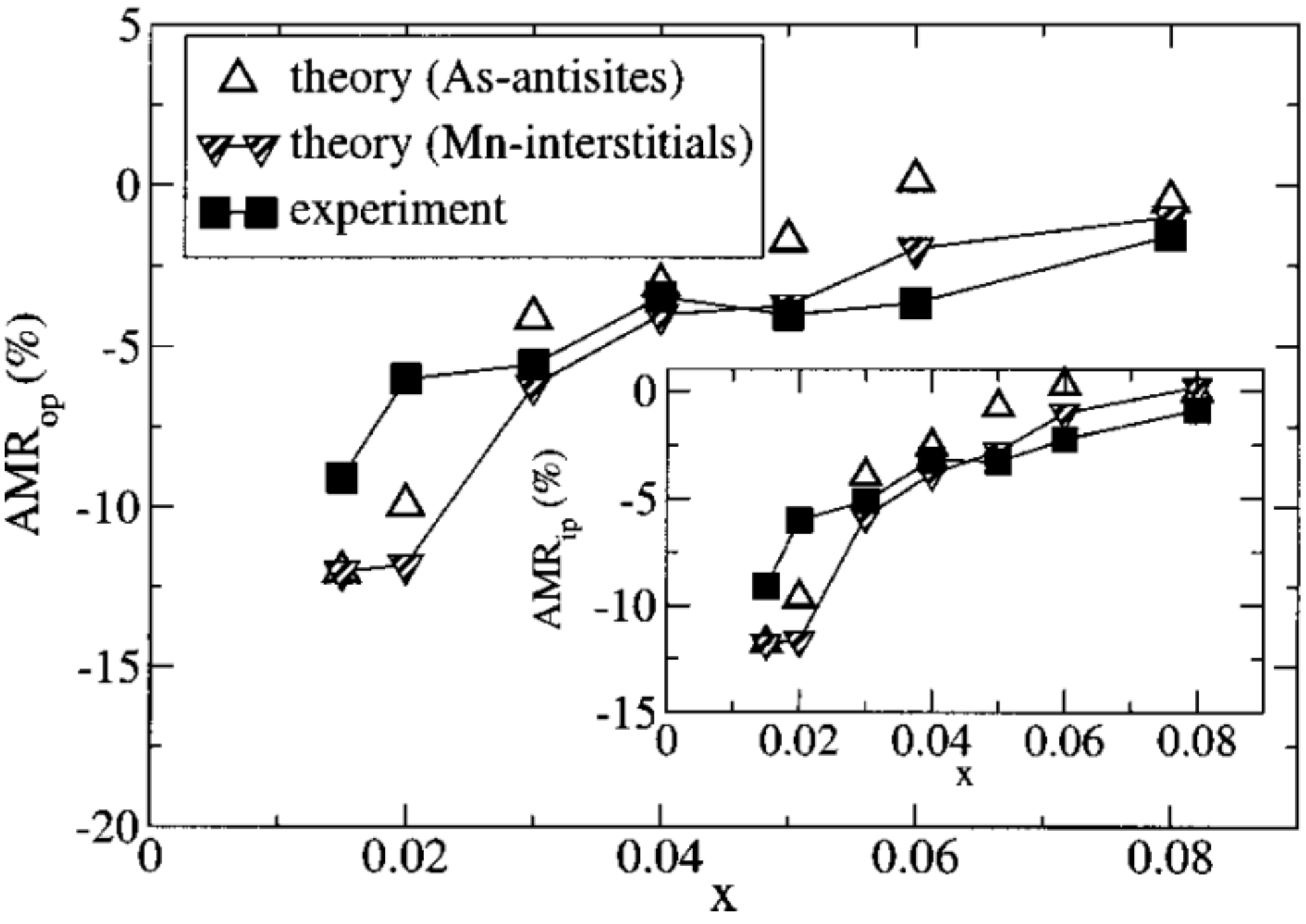}
  \includegraphics[scale=0.2,angle=0]{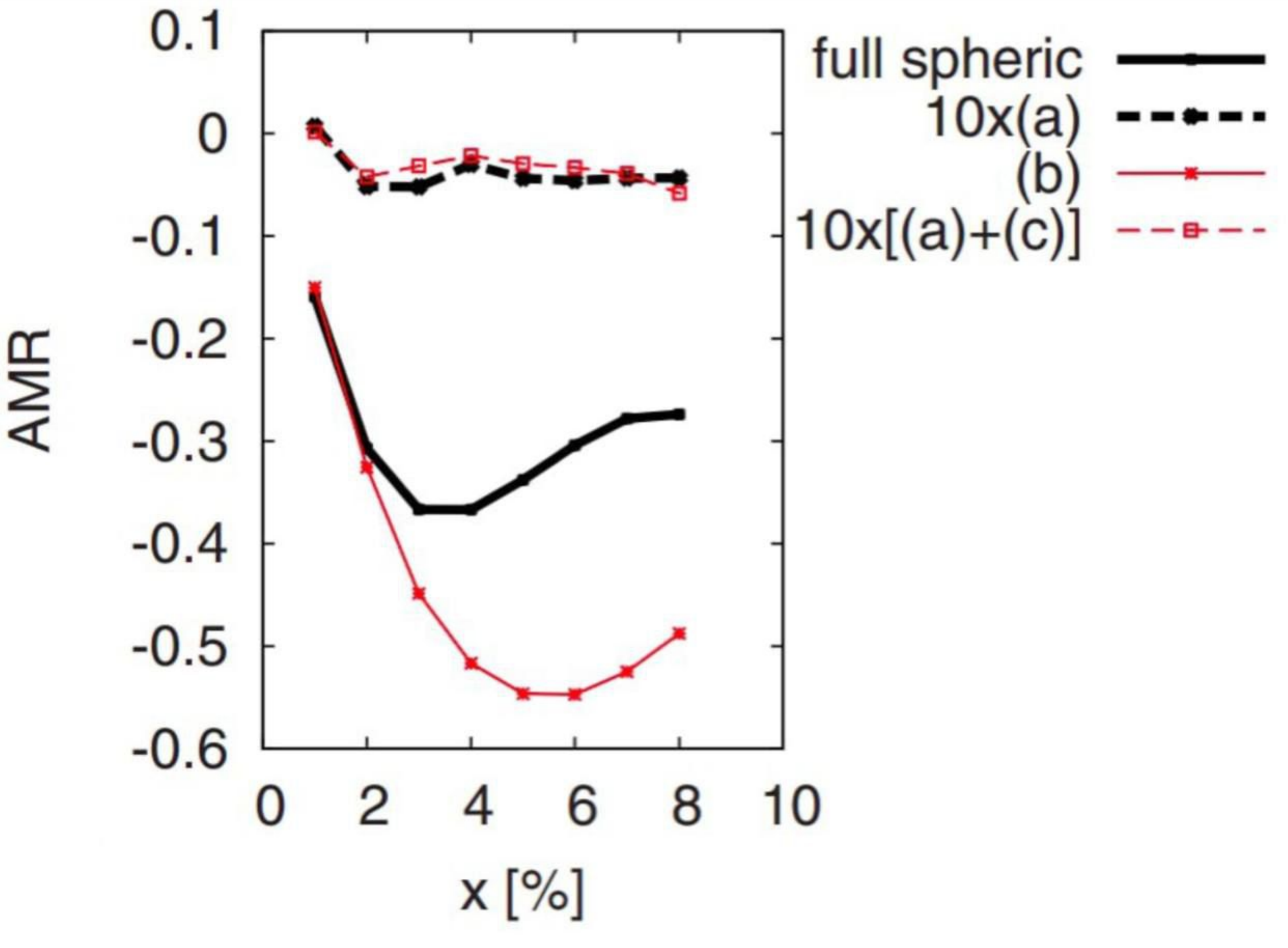}  
\caption{\lic{YES}{TJ+KV}{f11l}
  {\em Left:} Measured AMR in the dilute magnetic semiconductor
  (Ga,Mn)As with doping $x$ varied~\cite{Jungwirth:2003_a}. {\em Right:}
  Modelling allows to distinguish the intrinsic (a) and extrinsic (b,c)
  mechanisms of AMR; clearly, the extrinsic mechanism (b) as defined
  in Ref.~\cite{Vyborny:2009_a} dominates. Reproduced from (left) Ref.~\cite{Jungwirth:2003_a} and (right) Ref.~\cite{Vyborny:2009_a}.??}
      \label{fig-11}
\end{figure}

\subsubsection{Effective models}

Most transport phenomena depend on band structure solely in the vicinity
of Fermi level~\cite{note1} $E_F$. To that end, integral in Eq.~\ref{eq-06}
needs only limited knowledge of band structure (and Fermi velocity
component $v_x$); rather than using the band dispersion $E_{\vec{k}}$
in the full energy range, its effective model can often be constructed
which is easier to handle and offers better insight, e.g. into how the
magnetisation direction and spin-orbit interaction influence the band
anisotropy~\cite{Trushin:2009}. At this point, we remark that through
such anisotropy, the plasma frequency~(\ref{eq-06}) may become anisotropic:
in a non-magnetic cubic crystal, for example, $\omega_{p;xx}=\omega_{p;yy}$
but when magnetic order is present, $\vec{m}||\hat{x}$ breaks this symmetry. 
For the definition of such anisotropic $\omega_p$ and its discussion
related to intrinsic AMR, see Ref.~\cite{Nadvornik:2021_a}.

Turning our attention to the scattering, we first remark that should
the resistances in Eq.~\ref{eq-01} be calculated as $\propto 1/\sigma_0$
of~(\ref{eq-05}) for different directions of $\vec{m}$, whereas $\tau$ 
remains constant, the result is independent of $\tau$. In other words,
while scattering had to be taken into account to obtain finite conductivity
$\sigma_0$, it has no influence on the AMR. This is, however, only the
simplest situation possible: in most cases, $\tau$ does indeed depend on the
direction of $\vec{m}$ and this can either become manifest in the
matrix elements $M_{kk'}$ of the scattering operator (below, we give
an explicit example) or the direction cosine in Eq.~\ref{eq-07}. The
latter opens a pathway for the current direction to enter directly the
calculation of scattering time: $\tau$ in the relaxation time
approximation~\cite{Vyborny:2008_a} depends on $\vec{k}$ and the
Boltzmann expression for conductivity~\cite{Ahn:2021_a} assigns the largest
weight to $\tau(\vec{k})$ with $\vec{k}$ parallel to the current direction.

Such was the approach to understanding the AMR in elemental
ferromagnets (notably, nickel or iron) since the seminal work of
Smit~\cite{Smit:1951}. The ratio $\alpha$ of resistivities in
majority and minority spin channels (within what was later~\cite{Jaoul:1977_a}
called the {\em Smit mechanism}) allows to express the difference
$\Delta\rho=\rho_\parallel-\rho_\perp$ with respect to the direction of
$\vec{m}$ as
\begin{equation}\label{eq-10}
\frac{\Delta \rho}{\rho} = \gamma (\alpha - 1) 
\end{equation}
where $\gamma \approx 10^{-2}$ describes the competition of spin-orbit
interaction and exchange interaction.  For these simple cases it holds that
$\alpha$ > 1 (thus $\rho_\downarrow (T=0) > \rho_\uparrow (T=0) $), so that
the AMR is always positive and the other cases are described in
Fig.~\ref{fig-02}. It should be stressed that formula~\ref{eq-10} provides
only a basic guidance to AMR, yet it is referenced occasionally up to nowadays~\cite{Gomes:2019_a} when interpreting experiments; 
we return to the discussion of $sd$ models
applied to AMR in elemental metals and their alloys in Sec.~3\ref{sec_TM} and proceed 
now to discuss the effective models in dilute magnetic semiconductors (DMS).

While the previously discussed $sd$ models~\cite{McG-Potter} treat
the band structure only on a rudimentary level, essentially $\omega_p$
in~(\ref{eq-05}) is taken as coming from a single band and
independent of $\vec{m}$, models of transport in DMS are more
elaborate in this respect. The
valence band $E_{\vec{k}}$ can be obtained~\cite{Jungwirth:2006_a} from  
four- or six-band models (depending on the needed level of detail) and
conductivity can be evaluated using the Boltzmann equation, see Sec. 3\ref{sec_DMS}.
It turns out~\cite{Vyborny:2009_a} that the relaxation time
approximation (RTA) with a constant (magnetisation-direction independent)
$\tau$ leads to a too small AMR so that in the particular case of (Ga,Mn)As,
extrinsic mechanism (i.e. anisotropy of $\tau$) is dominant. The main
source of scattering, magnetic atoms (manganese) substituting for cations
of the host GaAs lattice, features magnetic and non-magnetic part
(their ratio is described by parameter $\alpha_{sc}$) and while analytical
estimates using Eq.~(\ref{eq-07}) such as
$$
  \mbox{AMR}=-\frac{20\alpha_{sc}^2-1}{24\alpha_{sc}^4-2\alpha_{sc}^2+1}
$$
can be obtained under simplifying assumptions, the full model shown in
Fig.~\ref{fig-11} reproduces the
measured~\cite{Jungwirth:2003_a} AMR well. Also, various combinations of
scattering and SO effects in two-dimensional electron gases have been explored:
extrinsic anisotropy in Dirac fermions~\cite{Trushin:2019_a} or
Rashba system~\cite{Kato:2007_a,Kato:2008_a,Trushin:2009}.

Turning our attention back to transition metals (see Tab.~II
in~\cite{McG-Potter} for a list of material systems), two important
publications should be mentioned. Mott \cite{Mott:1936} proposed that resistance in metals at high temperatures mainly depends on the scattering of 4s electrons into 3d states. At low temperature, the d-states are mainly populated, so that the main scattering is due to s-s-scattering and the resistivity is significantly lower.
Smit applied this idea first to AMR \cite{Smit:1951} and proposed that the AMR can be only due to spin-orbit interaction (i.e. neglecting the possibility of intrinsic AMR), should always be positiv and explained the larger AMR measured in dilute alloys by scattering due to foreign ferromagnetic atoms, where in simple transition metals (e.g. Ni) it is due to non-magnetic ions, lattice vibration or irregular stress. The foreign ferromagnetic atoms are supposed to have a larger effect on AMR than the other scattering effects, which also causes the AMR to decrease with increasing temperature (since lattice vibrations are becoming a more dominant contribution in resistance at higher temperatures) \cite{Smit:1951}.


\begin{figure}[!h]
  \begin{tabular}{cc}
    \includegraphics[scale=0.3,angle=0]{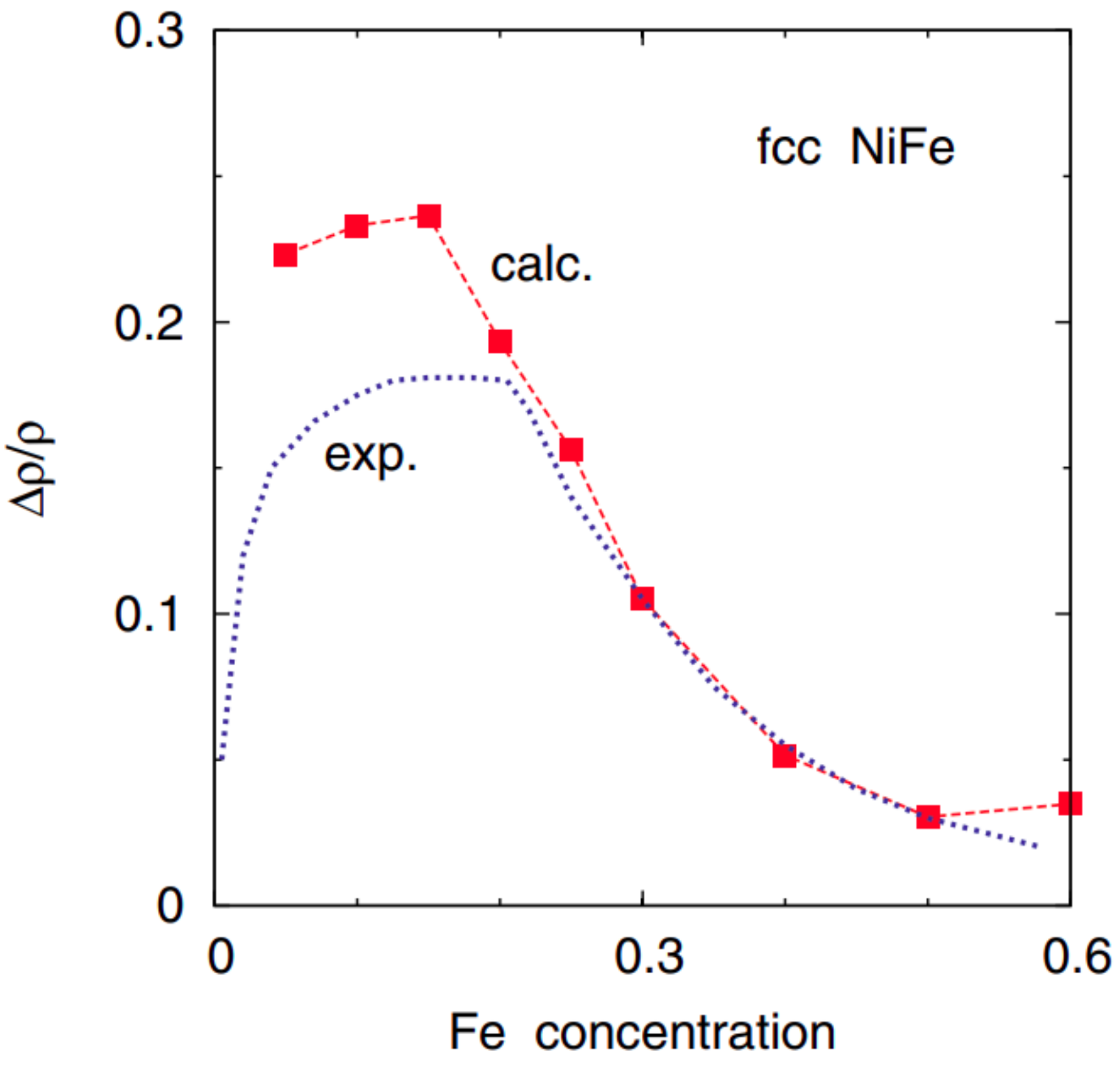} &
    \includegraphics[scale=0.25,angle=0]{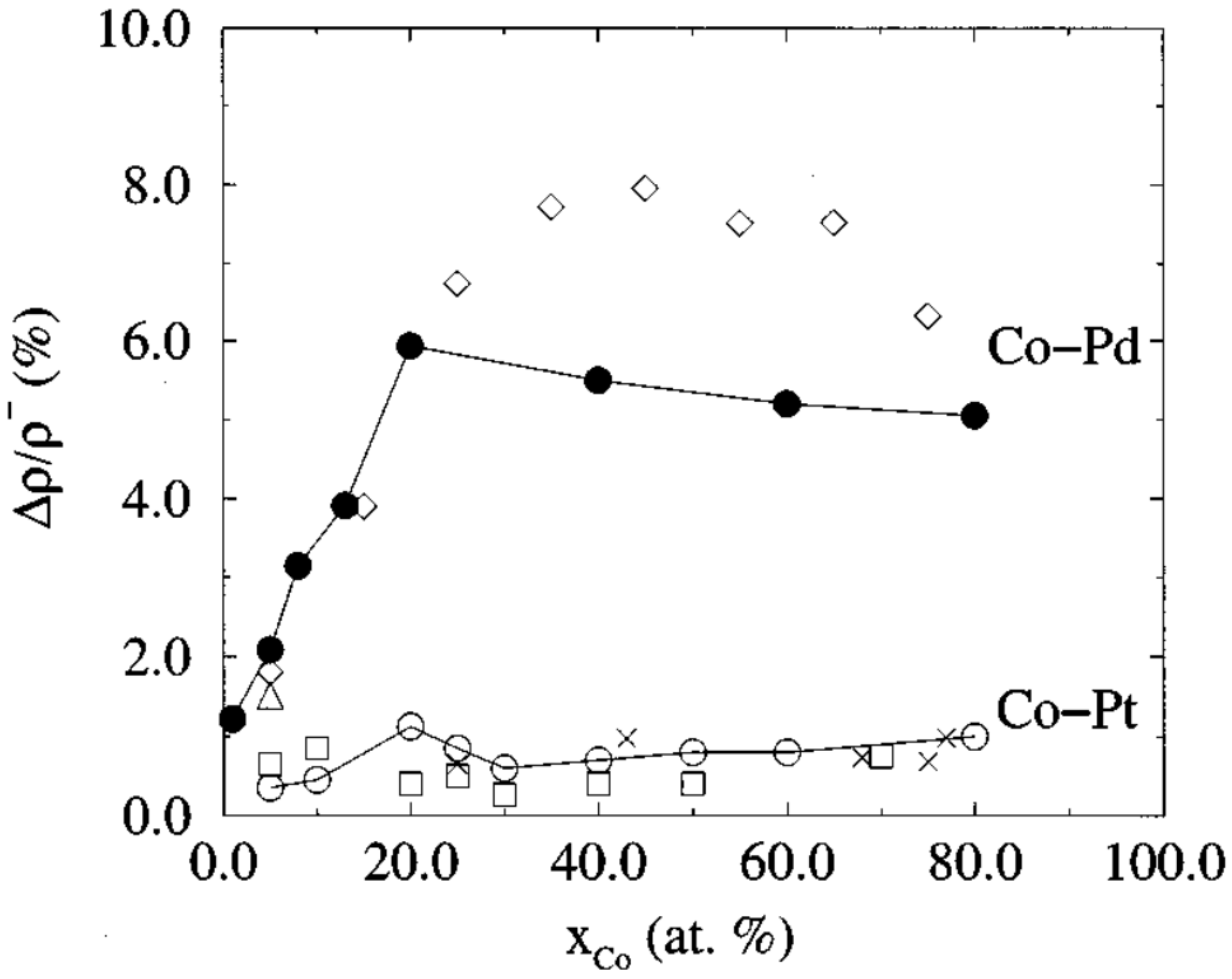} \\
    (a) & (b)
  \end{tabular}
\caption{\lic{YES}{Ilja+Hubert}{f12l}
  AMR in alloys (a) nickel-iron, (b) cobalt with non-mag. elements. Reproduced from (a) Ref.~\cite{Turek:2012_a} and (b) Ref.~\cite{Ebert:1996_a}.}
    \label{fig-12}
\end{figure}

\subsubsection{Ab initio models}
\label{sec_abinitio}

Most materials lack the simplicity of electronic structure which would
render construction of its effective model practicable. Band structure
can nevertheless be obtained by ab initio methods (DFT or beyond) and
should the AMR be dominated by intrinsic mechanism, plasma frequency
for different magnetisation directions can be calculated. Alternatively,
conductivity can be obtained using Green's functions 
$G=G^+(E_F)$ in Kubo formula~\cite{Banhart:1995_a}
\begin{equation}
  \sigma_{\mu\nu}(E)= \frac{e^2\hbar}{\pi V}
  \mbox{Tr } \langle v_\mu \mbox{Im }G v_\nu \mbox{Im }G\rangle
  \label{eq-08}
\end{equation}
%
by replacing the disorder average with $\tilde G v_\nu \tilde G v_\mu$
and $\tilde G^{-1}=E-H-i\Gamma$ with constant $\Gamma$
(which in the limit $\Gamma\to 0$ drops out from the expression
for AMR). When extrinsic mechanisms of AMR are important a better treatment
of scattering is needed and selfenergy $\Sigma$ (whereas Im~$\Sigma=\Gamma$)
must also be calculated by ab initio techniques.

The first attempt at such calculation has been undertaken
by Banhart and Ebert~\cite{Banhart:1995_a} who employed the coherent
potential approximation (CPA) but AMR as a function of $x$ (Fig.~1 in that
work) was overestimated. Further refinements were made~\cite{Ebert:2011_a}
and more recent calculations of Fe$_x$Ni$_{1-x}$ achieve a nearly
quantitative agreement~\cite{Turek:2012_a} to experimental AMR values.
A different approach, based on modelling the system by layers 
also reproduces well~\cite{Khmelevskyi:2003_a} the experimental data
on permalloy or Fe-Co~\cite{Freitas:1987_a} systems.
Temperature-dependent AMR has
now also been studied~\cite{Sipr:2020_a}. Beyond this material, cobalt
alloys (with Pt or Pd~\cite{Ebert:1996_a}) and nickel alloyed with Cu
or Cr~\cite{Vernes:2003_a} were studied, to give two examples among many.
AMR in permalloy doped by selected transition metals (see Fig.~2 in
Ref.~\cite{Sipr:2020_a}) agrees reasonably well with ab initio calculations
with the exception of doping by gold but it is presently unclear whether this
is a failure of CPA (in this particular case) or an experimental
issue~\cite{Yin:2015_a}. 
Recently, it has been argued (based on the same theoretical technique)
that in iron cobalt~\cite{Zeng:2020_a} the AMR is driven by intrinsic
mechanism.

\begin{figure}[!h]
\centering\includegraphics[scale=0.4,angle=0]{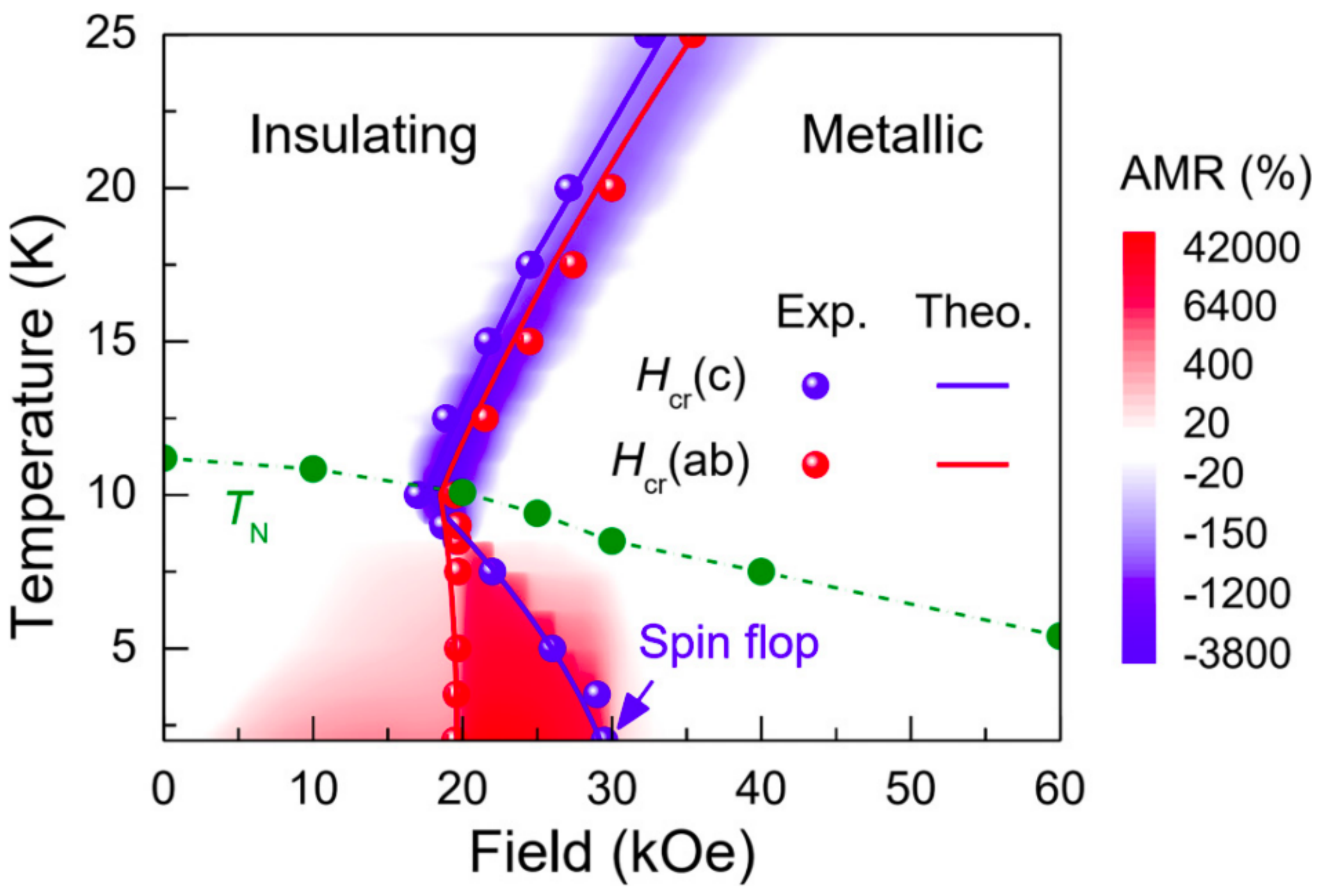}
\caption{\lic{YES}{Yang}{f04l}
   Magnitude of AMR in antiferromagnetic EuTe$_2$ where band structure
   changes from insulator to semi-metal depending on the configuration
	   of magnetic moments (which can be manipulated by applied magnetic field). Reproduced from Ref.~\cite{Yang:2021}.}
	\label{fig-04}  
\end{figure}

\subsection{Further remarks}

We conclude this section by several theoretical remarks before we proceed
to discussion of AMR in particular materials.

\textit{Hexagonal systems.}  In cubic systems, resistivity tensor reduces 
to a number $\rho_0$ (i.e. it is proportional to identity matrix); we will
now show that the same is true also for hexagonal systems.
Assume that $\hat{x}$ is parallel to one of the
sides of the hexagon. The two components of the resistivity tensor are
denoted as $\rho_{\parallel}$ and $\rho_{\perp}$ again. Then, if the
tensor is rotated by an angle $\theta$, its new form equals the form
presented in Eq. \ref{eq_rotrho}. In a hexagonal system, a rotation of
$\theta = \pi / 3$ is a symmetry operation and must not alter its
properties. In this case, the zero off-diagonal elements must be
conserved. In order to fulfull the equation $0 = \frac{1}{2} \Delta
\rho \sin(\frac{2 \pi}{3})$, we have to demand $\Delta \rho =
\rho_{\parallel} - \rho_{\perp} = 0$. Plugging this into
Eq. \ref{eq_rotrho}, the resulting resistivity tensor $\rho = \rho_0
\cdot \mathrm{I_2}$ \\

\textit{Metal-to-insulator transition.}  A very large change of electric
conductivity can be achieved by tuning the system between metallic and
insulating regimes: typical system being vanadium dioxide~\cite{Brito:2016_a}. 
While such typical MIT behaviour is unrelated to magnetism, proposals
of magnetic-order-dependent gap opening have appeared for orthorhombic
CuMnAs~\cite{Smejkal:2017_a} and experimentally, semimetallic antiferromagnet
EuTe$_2$ discussed in Sec. 3(c) is the first system where the transition
between low- and high-resistance states was achieved~\cite{Yang:2021} by
rotating the magnetic moments as the phase diagram in Fig.~\ref{fig-04} shows.
This effect can be understood as the extreme case of intrinsic AMR: rather
than deforming Fermi surface (FS) slightly by rotating the magnetic moments,
the FS disappears altogether. A related effect can also occur
in magnetic topological insulators, see Sec.4b
~\ref{exoticstuff}.\\

\textit{Relative and absolute AMR.} It is customary to evaluate the
AMR in relative terms. This makes good sense for extrinsic AMR
where both $\rho_0$ and $\Delta\rho_0$ are proportional to the density of
scatterers $n_{imp}$ and the ratio~(\ref{eq-01}) is then independent of
$n_{imp}$. Fig.~4 in~\cite{Isnaini:2020_a} demonstrates that this may
be true for a large group of samples. On the other hand, when resistivity
comprises two additive parts (in the spirit of Matthiessen's rule) where
one is anisotropic and the other is not, it is more meaningful to focus
on absolute difference of resistivities for magnetic moments parallel and
perpendicular to current. This is also the case for polycrystalline samples
where the isotropic part of resistivity is due to scattering on grain
boundaries: a suitable approach is then the Fuchs-Sondheimer theory
discussed e.g. in the introduction of Ref.~\cite{Rijks:1995_a}.

\section{Materials}
\label{ch_Mat}


\subsection{Elemental ferromagnets (TMs) and their alloys}
\label{sec_TM}

First observation of AMR was made in iron and nickel with cobalt following
(see Sec.~1\ref{sec_History} for the history) and understandably, the first microscopic theories
therefore aimed at elemental ferromagnets. The first step beyond the
quantification of the AMR ratio on the level of Eq.~\ref{eq-01} was to analyse
individual symmetry contributions to the AMR~\cite{Doring:1938} (as given by
Eq.~\ref{F_Prague} or Eq.~\ref{eq_long010C1} nowadays) and next, their 
temperature dependence was determined~\cite{vanGorkom:2001_a,Kimling:2013_a}.
Other papers on these materials are discussed in
Sec.~2(b)\ref{sec_abinitio} since the results of them are outdated by
now, but of great historical importance in the development of
models. In the following, the most interesting results are discussed
and at the end of this section, typical AMR values are listed in table
\ref{table-TM}. There is some scatter in the values of AMR, whose
origin cannot be conclusively identified, since part of the
information is lacking in some of the studies. So it is unclear
whether in all studies the saturation magnetization is reached, what
is the crystal structure and crystalline quality and in very thin
films, surface scattering can even play a role. Important observables
to watch are the sign of the AMR and the order of magnitude of the values.

State-of-the-art reports of AMR in the three transition metals are for
iron thin layers by van~Gorkom~et~al.~\cite{vanGorkom:2001_a} and for nickel films
by Xiao~et~al.~\cite{Xiao:2015_b}. While these two metals
are cubic, the situation is somewhat more complex for cobalt which 
exists in the hcp~\cite{Bakonyi:2002_a} and fcc~\cite{Xiao:2015_a}
phases. In polycrystalline samples~\cite{El-Tahawy:2022}, the AMR is
a factor of about 1.8 larger for fcc than for hcp (hexagonal close
packing). This behavior was explained by differences in
(calculated) DOS at the Fermi level. The hcp-Co AMR is reported to lie
between 1.14 and 1.23 \% and for predominantly fcc-Co samples the span
is 1.73 to 2.19\%.

Polycrystalline Co has a dominant intrinsic AMR contribution, which was shown by frequency-dependent studies \cite{Nadvornik:2021_a} on ac-AMR (See Sec.~4\ref{sec_ApplFurther} for details on the ac-AMR method). In the same study, it was also shown experimentally that polycrystalline Ni and the alloys Ni$_x$Fe$_{(100-x)}$ with x = 50 and x = 81 (permalloy) have a negligible intrinsic contribution and plasma frequency calculations indicate a similar behavior in single-crystalline materials. For the single-crystalline case an experimental confirmation is still required. For iron, such investigations are lacking entirely and it is thus unknown whether the AMR in Fe is caused by extrinsic or intrinsic contributions.


In the analysis of D\"oring \cite{Doring:1938} based on
Eq.~\ref{F_Prague} or Eq.~\ref{eq_long010C1}, four-fold signals were
also identified in single-crystalline nickel. However the reporting of
higher-order signals in these basic TMs is rare \cite{Xiao:2015_b} and
usually only two-fold signals are reported. In some studies,
deviations from two-fold AMR are accounted for by magnetocrystalline
anisotropy using Stoner-Wohlfahrth approaches as for example was
reported by Miao et al. in single-crystalline Co and polycrystalline
Fe$_{20}$Ni$_{80}$~\cite{Miao:2020} as well as in epitaxial
Fe$_{30}$Co$_{70}$ thin-films~\cite{Miao:2021}. 

Alloys offer a vast field for research on AMR, since the effect can be
increased significantly, as compared to the pure TMs, by tuning their
composition. Our discussion of alloys is split into two categories:
first, the three basic TMs with small amounts of TM impurities are discussed
and second, we focus on alloys made from a combination of the three
basic TMs. The best known example of the second category is Permalloy
(Ni$_{80}$Fe$_{20}$). Typical values of AMR ratios are listed for the
alloys in table~\ref{table-TM} as well. 

A comprehensive work on the first category of alloys, nickel with TM
impurities, is Jaoul \textit{et al.}~\cite{Jaoul:1977_a}. An important
characteristics of these impurities is the virtual bound state (VBS);
when the VBS appears~\cite{Levy:2016_a} (for example with V, Cr, Os or Ir)
both positive and negative AMR was measured and otherwise the AMR remains
positive (this was the case with Mn, Fe, Co, Pd, Cu, Zn, Al, Si, Sn and Au
where the VBS does not appear).
This was attributed to
the effect of the $L_z S_z$ operator of the spin-orbit interaction on
the VBS, which was included into the description of AMR by adding the term $+ 3 \beta\alpha/(\alpha+1)$ to formula~\ref{eq-10}, where $\beta$ encrypts the effect of the $L_z S_z$ term. It can be positive or negative, thus the AMR can show both signs. Please note, that this explanation for negative AMR is consistent with the more recent and elaborate one given by Kokado and Tsunoda \cite{Kokado:2012} (see sec. \ref{sec_History}). In contrary to the latter ones, the extension of Eq.~\ref{eq-10} by Jaoul is limited to strong ferromagnets and is not capable of describing e.g. features of half-metals such as spin-dependent effective mass.
In another study by McGuire et al.~\cite{McGuire:1984}, robust
negative AMR up to room temperature (RT) was achieved by considering
Ir as an impurity in various hosts such as nickel, cobalt, iron and in
certain alloys of these three.


%
%

In the second category of alloys we find the combinations FeCo, CoNi
and NiFe as well as FeCoNi in diverse compositions. AMR in these
alloys is robust and typically one order of magnitude larger than in the pure TM as can be seen in table~\ref{table-TM}. Many publications focus on AMR measurements for different compositions and track the dependency of AMR on concentration of a certain element. Of special interest is permalloy, which shows not only a large AMR, but is also used in a number of industrial applications, for example in magnetic readout heads. The interest of industry is due to its nearly zero magnetostriction and high magnetic permeability.

Composition-dependent studies of the AMR ratio in the nickel-based alloys Fe$_x$Ni$_{1-x}$, Co$_x$Ni$_{1-x}$ and (Co$_x$Ni$_{1-x}$)$_{86}$Fe$_{14}$ are provided by Ishio \textit{et al.} \cite{Ishio:1998_a} and in the iron-based alloys NiFe and FeCo by Berger \textit{et al.} \cite{Berger:1988_a}. In the first case of the nickel-based alloys Ishio \textit{et al.} investigated the AMR ratio for two different current directions [001] (which they call $K_1$) and [111] (which they call $K_2$). Extremal values of AMR are achieved for (Ni$_{80}$Co$_{20}$)$_{86}$Fe$_{14}$ with $K_1 = +68 \%$ and $K_2 = -32 \%$. For $K_2$ there is an increase leading to a sign change to positive values with increasing Fe and Co \cite{Ishio:1998_a}. This is consistent with other measurements reporting AMR of up to 50$\%$ in NiCoFe alloys with a maximum at Ni$_{80}$Co$_{20}$Fe$_{5}$ \cite{Miyazaki:1990}.
In the FeNi alloys, a maximum AMR ($K_1$) of ca. 35$\%$ is achieved at ca. 10-15 $\%$ Fe. Permalloy shows an AMR of 25$\%$ \cite{Ishio:1998_a}. In the second case of the iron-based alloys by Berger \textit{et al.}, the AMR is split into an impurity-based AMR contribution $(\frac{\Delta \rho}{\rho})_{im}$ and a phonon-based contribution $(\frac{\Delta \rho}{\rho})_{ph}$. Both contributions are indiviually plotted vs. the iron concentration. $(\frac{\Delta \rho}{\rho})_{ph}$ is positive for the case of weak electron scattering in Fe-Co and negative in the case of strong, resonant electron scattering in the other alloys. The impurity contribution is always positive and larger for the strong scattering. A peak AMR of ca. $16\%$ is found for permalloy. 
A more recent study~\cite{Miao:2020} reports only few per cent AMR 
for sputtered Ni$_{80}$Fe$_{20}$ films but in absolute terms, i.e.
$R_\parallel-R_\perp$ in Eq.~\ref{eq-01}, the anisotropy is similar
in both samples; here, the buffer layer thickness also plays role~\cite{w-todo}
most likely through changing the background resistivity.
AMR in epitaxial Fe$_{30}$Co$_{70}$ was shown to have strong crystalline
terms~\cite{Miao:2021}. More alloys involving transition metals are
discussed in Sec.~3(f).


A comparison of experimental data and CPA calculations is given in
Fig.~\ref{fig-12}(a), where the AMR ratio is calculated for fcc NiFe
alloys dependent on the Fe concentration. Especially for
concentrations larger than 0.15, the calculations describe the
experimental data almost perfectly. Calculations of dilute
\underline{Ni}Fe thin-film, wires and FM/non-magnetic/FM multilayers
using Boltzmann equation with RTA and a two-current model are carried
out by Rijks \textit{et al.} \cite{Rijks:1995_a}. 



\begin{table}
\begin{tabular}{ccl}
	Material & AMR percentage & Remarks \\
	\hline
	Fe & 0.2 - 1.5 & RT~\cite{Dahlberg:1988_a} to
                         low temperature (LT)~\cite{vanGorkom:2001_a} \\
	Ni & 1.8 - 3.15 & see Tab.~I in\cite{Isnaini:2020_a} and also Fig.~\ref{fig-01}a\\
	Co & 0.3 - 3.5 & from \cite{El-Tahawy:2022}; Fig.1 in~\cite{Bates:1946_a}
	  \\
	\hline
        Ni \mbox{ with }Pd & 2 & $T = 4.2 $ K; impurity without VBS \cite{Jaoul:1977_a}\\
        Ni \mbox{ with }Zn & 6.5 & low temperature; impurity without VBS \cite{Jaoul:1977_a} \\   
        Ni \mbox{ with }Cr  & -0.28 & $T = 4.2 $ K; impurity with VBS \cite{Jaoul:1977_a}  \\     
... & & \\
		Co \mbox { with } 3\% Ir & -2.56 & RT; \cite{McGuire:1984} \\
... & & \\
Co$_{45}$Pd$_{55}$  & 7.96 & $T = 4$ K; \cite{Jen:1992_a} \\
... & & \\
	\hline
	Ni$_{80}$Fe$_{20}$ (permalloy) & 16 - 25 & LT and RT \cite{Berger:1988_a} ; $T = 10$ K \cite{Ishio:1998_a} \\
	(Ni$_{80}$Co$_{20}$)$_{86}$Fe$_{14}$ & +68 and -32& [001] and [111] current direction, resp. \cite{Ishio:1998_a} \\
	Ni$_{77}$Fe$_{22}$Cr$_2$ & 0.76 & $T = 4.2$ K; \cite{Chakraborty:1998_a} \\
	... & & \\
	
\end{tabular}
\caption{Examples of AMR values for three groups of TM-based systems: pure
room temperature (RT)-FM metals (i.e. Fe, Co, Ni); the basic TM with TM-impurity; and alloys of the three basic TM. More examples of Ni-alloys with other TM impurities can be found in table 1 of \cite{Jaoul:1977_a} and more example of alloys with Ir as impurity are listed in table 1 of \cite{McGuire:1984}. AMR for other concentrations of Pd in CoPd are listed in table 1 of \cite{Jen:1992_a}, where the given composition Co$_{45}$Pd$_{55}$ shows the maximum value.
A broader listing of the nickel-based alloys FeNi, CoNi and (CoNi)Fe is found in \cite{Ishio:1998_a} and of the iron-based alloys NiFe, FeCr, FeV and FeCo in Fig. 1b and Fig. 2 of \cite{Berger:1988_a}. AMR values for NiFeCr with higher concentrations of Cr are listed in Tab. 1 of \cite{Chakraborty:1998_a}.}
\label{table-TM}
\end{table}



\subsection{Dilute Magnetic Semiconductors}
\label{sec_DMS}

A completely different perspective of AMR is offered by
the dilute magnetic semiconductors (DMSs): magnetism and transport
properties can be tuned in these systems to some extent independently.
Our understanding of the electronic structure in DMSs relies on the
solid knowledge about III-V (and other) systems such as GaAs combined
with substitutional effect of a magnetic element (typically manganese)
whereas coupling between localised magnetic moments (provided in that
case by 3$d^5$ electrons) is mediated by delocalised
carriers~\cite{Jungwirth:2006_a}. The key parameter is the acceptor
(in case of III-V:Mn) binding energy $E_0$ 
and also its physical origin~\cite{Masek:2010_a} indirectly influences
the magnetotransport mechanism. 


Given the appreciable spin-orbit interaction in GaAs ($\Delta_{SO}=0.34$~eV)
and basically metallic conduction (Fig.~32 in~\cite{Jungwirth:2006_a}),
AMR could have been anticipated to occur in (Ga,Mn)As. Indeed, the
first report of AMR in (Ga,Mn)As~\cite{Baxter:2002_a} has soon
been followed by more detailed studies~\cite{Jungwirth:2003_a,Wang:2005_a}
and new ideas keep appearing (co-doping by lithium~\cite{Miyakozawa:2016_a}
or As/Sb substitution~\cite{Howells:2013_a,Wang:2019_a}). These studies 
allowed to explore the AMR under continuous variation of band-structure
parameters and filling as well as of strain~\cite{Limmer:2006,Limmer:2008}.

\begin{figure}[h]
  \includegraphics[angle=0,scale=0.6]{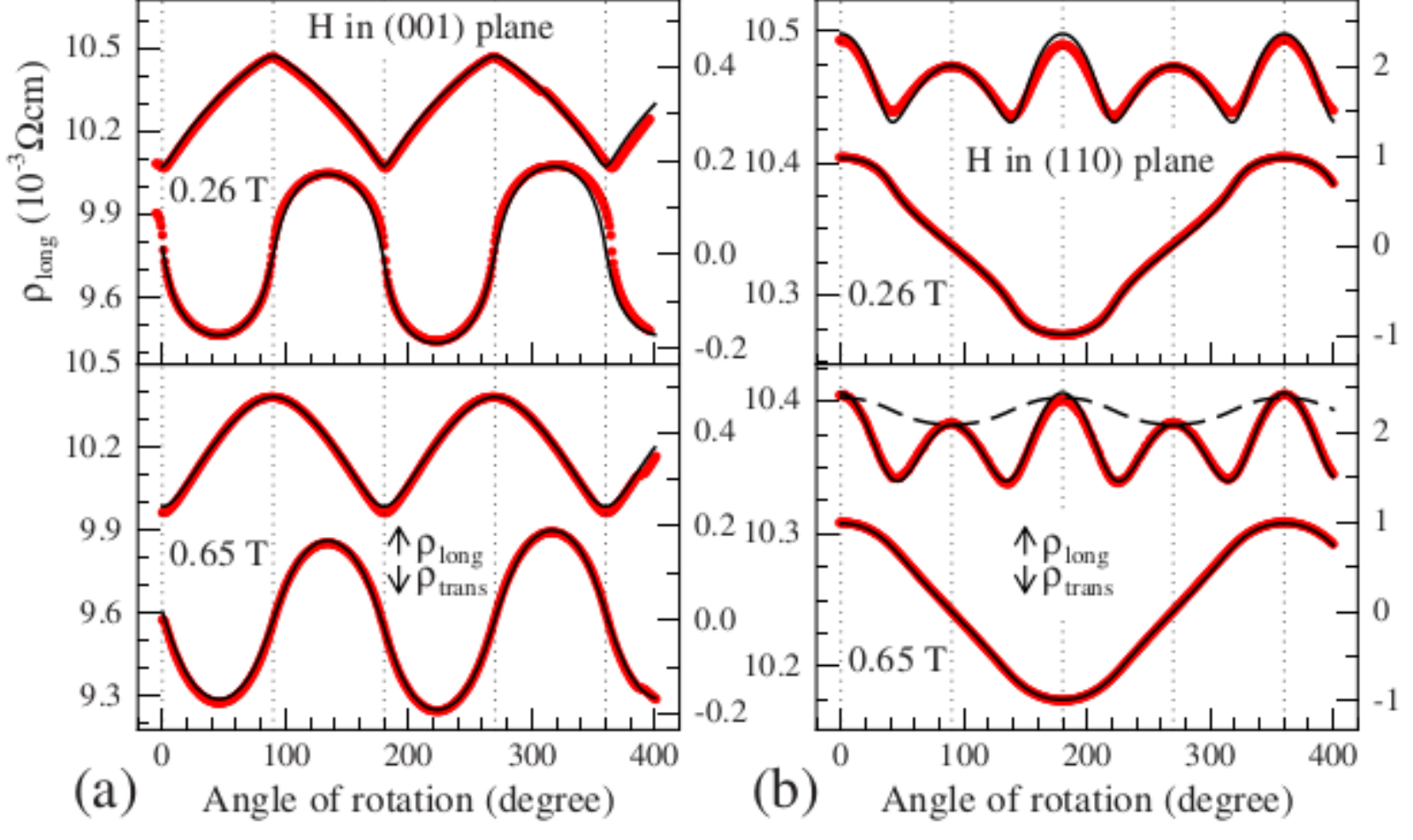}
\caption{\lic{YES}{STBG}{f10l}
    The data (red thick lines) and fit (black thin line) of the longitudinal resistivity $\rho_{long}$ (upper line in every plot) for current direction along [110]. A magnetic field of 0.26 and 0.65 T are rotated in the (001) and (110) plane, respectively. The dashed line in the lower panel of b) refers to an attempt of fitting the data to $\cos(2\phi)$, which is clearly insufficient. The lower lines in every plot are the transversal resistivity. Reproduced Detail from Fig. 7 in Ref.~\cite{Limmer:2008}.}
  \label{fig-10}
\end{figure}

Research on AMR in DMSs has pushed the understanding from 'complicated
to simple' concepts: idealised $sd$ models~\cite{McG-Potter} gave way
to a semiquantitative description~\cite{Jungwirth:2002_a} where the
intrinsic and extrinsic sources of AMR~\cite{Nadvornik:2021_a} could
be separated (see Sec. \ref{sec_notAMR} and then the detailed discussion of
microscopic models in Sec. \ref{ch_Model}). An application of this model is shown in Fig. \ref{fig-10}, where a Stoner-Wohlfarth model as well as four-fold crystalline AMR terms have to be taken into account. In the lower panel of b) it can be clearly seen that an attempt of fitting the data to terms without four-fold terms leads to insufficent agreement. Currently, interest in the once very
popular (Ga,Mn)As subsides since the prospects for the RT
magnetism~\cite{Jungwirth:2005_a} remain unfulfilled. Nevertheless,
Mn-doped III-V semiconductors remain a good test-bed for exploring
transport phenomena in materials with tunable magnetic properties.

Despite the versatility of this material class, not much attention was
given to other DMS: 
two-fold and eight-fold AMR were reported in a 10 nm film of (In,Fe)As
in \cite{PhanNamHai:2012_a}. In a 100 nm film of the same material,
the eight-fold component was missing, which was attributed to higher
electron concentration. Yet this claim is not supported by microscopic 
calculations and, together with the very small magnitude of AMR and low
electron concetrations, this may well be a hint that it is not the absence of
$sd$-scattering at the Fermi surface~\cite{PhanNamHai:2012_a} but issues
with sample quality that lead to this unusual behaviour. Better established
materials, in terms of sample quality, such as (Cd,Mn)Te still suffer from
too low carrier concentration~\cite{Terry:1996_a} and even if the regime of
metallic conduction is reached~\cite{Shapira:1990_a}, only magnetoresistance
(rather than its anisotropy) is measured and transport mechanisms seem to
be less well-established than in the case of (Ga,Mn)As. These systems
also occasionally suffer from the formation of multiple phases~\cite{ANQ2019}.
Finally, magnetically doped A$_2$B$_3$ systems (where A is either Bi
or Sb and B is Se or Te) should be mentioned~\cite{Lee:2014_a}
or~\cite{Dyck:2005_a}. Finally, magnetically doped ZnO should be
mentioned~\cite{Lee:2006_a} 
whereas the mechanism of magnetic state formation is complicated and
the latter can even be achieved by hydrogenation of ZnO~\cite{Khalid:2012_a}.





\subsection{Antiferromagnets}
\label{sec_AFM}
While ferromagnetism has been a known phenomenon since ancient times, its
counterpart antiferromagnetism was introduced no earlier than in 1933
by Landau \cite{Landau:1933}. It is little surprising then, that AMR
in this material class has only recently been investigated. A little
less than ten years ago, first studies appeared reporting AMR in
antiferromagnetic (AFM) Sr$_2$IrO$_4$ \cite{Fina:2014_a, Wang:2014_a}
and in FeRh which undergoes a transition from antiferromagnet to FM
\cite{Marti:2014_a,Turek15}. In recent years, the class of AFM
materials has received more attention due to the development of AFM
spintronics~\cite{Baltz-AMRrev}. The hope is to revolutionize
spintronic applications by making use of advantageous properties of
AFMs such as robustness against magnetic field perturbations, the lack
of a stray field or ultrafast dynamics. A prototypical magnetic memory
was developed using CuMnAs, see Sec. 4(c), and the transversal component of AMR
(also called the planar Hall effect), was used as readout \cite{Wadley:2016}.
%
%
As tetragonal CuMnAs has thus become a popular AFM material~\cite{Zubac:2021_a},
its properties came under intense scrutiny; microscopic mechanism of its AMR
is however far from clear~\cite{Volny:2020_a}: multiple kinds of impurities
lead to AMR which is comparable to experiment. On the other hand,
intrinsic AMR linked to gap opening controled by N\'eel vector orientation
was proposed\cite{Smejkal:2017_a} to occur for the orthorhombic
phase\cite{Emmanouilidou:2017_a} of CuMnAs which is similar\cite{Bodnar:2018_a}
to another AFM metal: Mn$_2$Au.

\begin{figure}[h]
  \includegraphics[angle=0,scale=0.6]{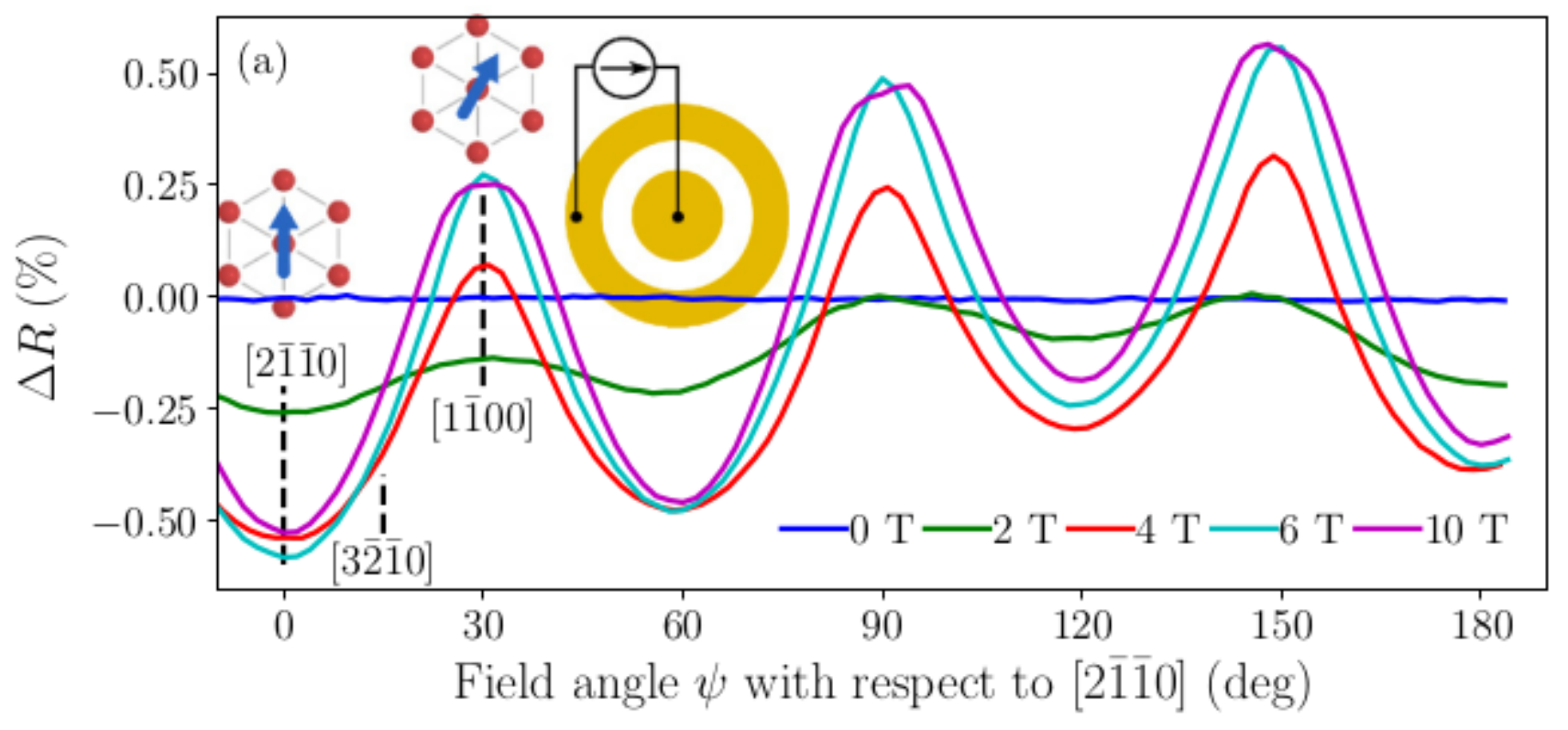}
\caption{\lic{YES}{us}{f07l}
    Crystalline AMR measured in MnTe for different field strengths using a Corbino geometry. The AMR shows a six-fold symmetry, which can be expected for the crystalline AMR in a hexagonal material. Reproduced from Ref.~\cite{Kriegner:2017_a}.}
  \label{fig-07}
\end{figure}

Another material which is a candidate for magnetic memory is MnTe: Next to a robust, continously varying AMR signal suitable as readout for AFM states, stability of the AFM states against pertubing magnetic field itself was shown by means of zero-field-AMR (zf-AMR)~\cite{Kriegner:2016_a}: resistivity is measured in zero magnetic field at low temperatures after the sample is field-cooled in a \textit{writing field}. After taking a data point, the sample is heated up again and the procedure repeated for another orientation of the writing field. Repeating this for a continous rotation of writing fields yielding a periodically zf-AMR signal resembling the conventional AMR. Furthermore, in the experiment it was shown that for a writing field of 2 T, the zf-AMR is multistable against pertubations from magnetic fields of 1 T or less. Hence, the possibility of writing and readout combined with robustness against pertubing fields makes it an excellent candidate for a spintronic device~\cite{Kriegner:2016_a}.
Also, crystalline AMR measured in the Corbino geometry shown in Fig.~\ref{fig-07} shows a strong $\cos 6\varphi$ component due to the hexagonal crystalline structure of MnTe~\cite{Kriegner:2017_a}.\\
Point-contact measurements in single-crystalline bulk sample of the AFM Mott-insulator Sr$_2$IrO$_4$ at liquid nitrogen temperatue yielded a field-dependent transition from four-fold AMR (low field) to two-fold AMR (high field). The four-fold AMR was interpreted as crystalline AMR reflecting the tetragonal crystal structure of the single-crystalline sample, while the transition to two-fold AMR being due to canting of AFM moments. The AMR ratio shows a maximum of $14 \%$ at a field of 120 mT. The large AMR has been attributed to large SOC in this 5d oxide \cite{Wang:2014_a}. 
In another experiment, AMR in a Sr$_2$IrO$_4$ film is studied by utilizing a SIO/La$_{2/3}$Sr$_{1/3}$MnO$_3$ (LSMO) heterostructure. The ferromagnetic LSMO is used to control the reorientation of AFM spin-axis via exchange spring effect. The AMR at low temperatures ($T = 4.2$ K) is showing a 4-fold behavior, while at intermediate temperatures ($T = 40$ K) no AMR signal was detected and at higher temperatures ($T=200$ K) the AMR is dominanted by the two-fold AMR of the FM LSMO~\cite{Fina:2014_a}. \\
An AFM memory in FeRh was proposed by Marti \textit{et al.} where field-cooling is used to write a magnetic state and AMR used as readout. Similarily to MnTe \cite{Kriegner:2016_a}, 
the memory shows a certain insensitivity against pertubing fields~\cite{Marti:2014_a}. 
RhFe undergoes a FM-AFM transition. It is antiferromagnetic below $T_N = 370$ K and ferromagnetic between $T_N$ and $T_C = 670$ K. Transport was investigated for both phases by means of first-principle calculations (relativistic TB-LMTO). AMR exists in both the FM and the AFM phase and was stated to be in a range of up to 2\% depending on the Rh-content. The AMR in the AFM phase is larger for most of the investigated compositions \cite{Turek15}.


A rather special case are single-crystals of AFM EuTe$_2$ where a peak value of 40000 \% at 2 K and 22 kOe (2.2 T) is achieved due to a metal-insulator phase transition (MIT). Since the MIT shows different critical fields for the ab-plane and the c-axis, the AMR becomes collosal for applied field values between the in-plane and the out-of-plane critical fields leading to a several order of magnitude change in resistivity for rotating the magnetic field. Bandstructure calculations confirmed this behavior.
AMR for fields and temperatures entirely within one phase (metallic or insulating) is in an order of magnitude of less than 20$\%$ and thus comparable to other materials \cite{Yang:2021}.
Finally, we remark that (collinear) ferrimagnets can be
considered to belong to this group too~\cite{Kabara:2014}, since they
equivalently have two magnetic sublattices, with the difference that
the magnetic moment is not fully compensated. Furthermore, AMR in
non-collinear AFMs is a rather novel topic and will be discussed in
sec. 4b-\ref{ssec_ncoll}.




\subsection{Heusler alloys}
\label{sec_Heusler}

\textit{Introduction}. Heusler compounds exhibit a large variety of
fascinating properties, as for example ferromagnetism and
antiferromagnetism, thermoelectricity, high spin-polarization,
superconductivity and topological features \cite{Manna:2018}. In
general, their formula is X$_2$YZ, where X and Y are transition metals
and Z is a main group element. X is more electropositive than Y. If X
and Y are exchanged, the material is called an inverse-Heusler. There
are so-called half-Heuslers, which are given by the formula XYZ
\cite{Manna:2018}. In general, Heusler compounds have a cubic crystal
structure, which can occur in different variations. The first Heusler
compound was Cu$_2$MnSn, discovered already in 1903, which was a
surprise because it was ferromagnetic while its components are
not~\cite{Heusler:1903}.

Despite the generality of its definition, a large body of research is
focused on cobalt-based Heusler alloys (thus Co$_2$YZ and Y is typically
Mn, Fe or a lighter 3d element), since they generally show important
features interesting for potential spintronics applications, such as
relatively high Curie temperatures, half-metallicity, large
magnetotransport effects and many more.  

\begin{figure}[h]
  \includegraphics[angle=0,scale=0.5]{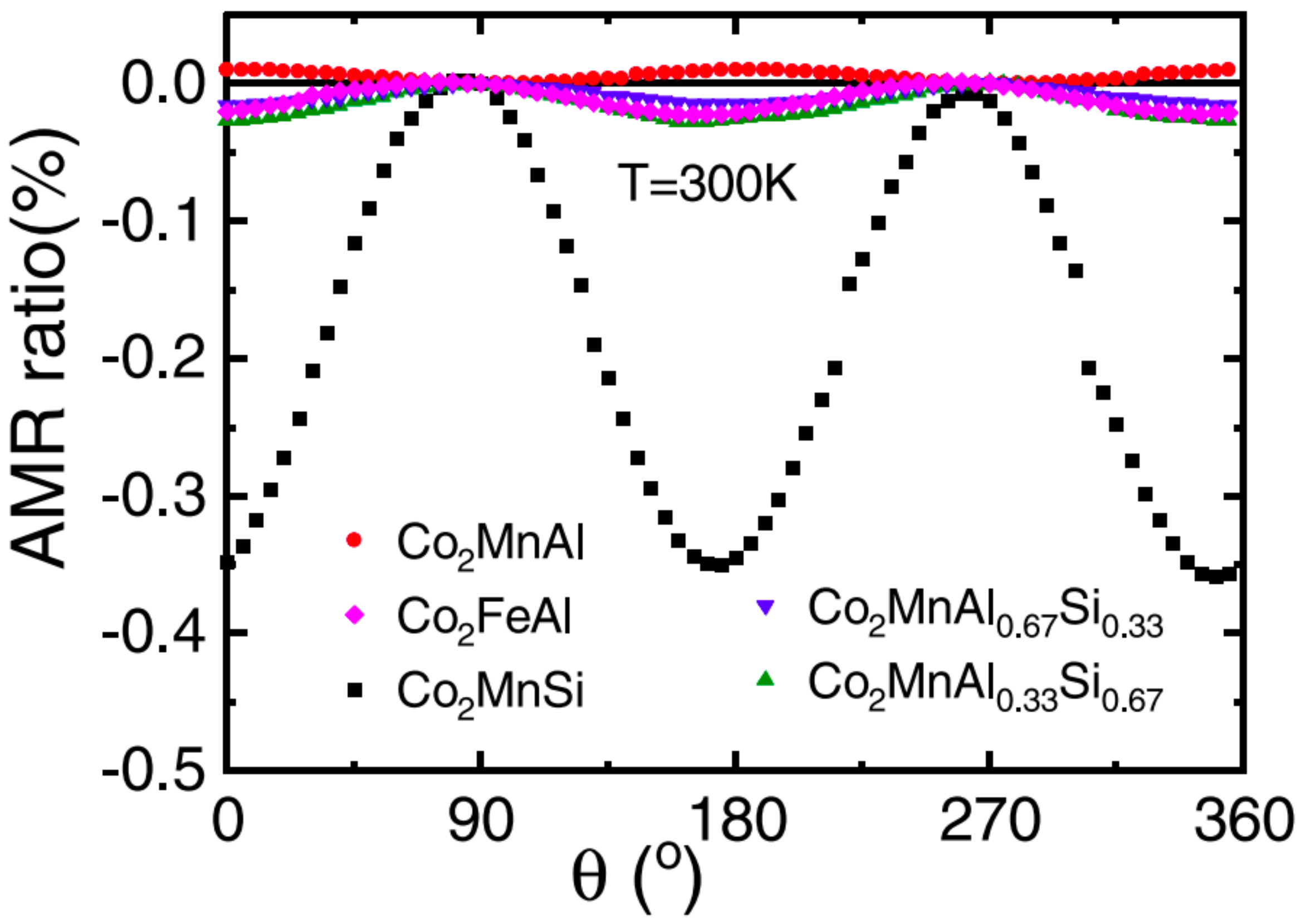}
\caption{\lic{YES}{Crowell}{f06l}
    AMR in Co$_2$MnAl (CMA), Co$_2$FeAl, Co$_2$MnSi (CMS) and Co$_2$MnAl$_x$Si$_{1-x}$ for x = 0.33 and 0.67. The AMR for CMS has the largest magnitude. All materials exhibit negative AMR except apart from CMA. The order of magnitude of the AMR is in agreement with other studies of Co-based Heusler alloys. Reproduced from Ref.~\cite{Breidenbach:2022_a}.}
  \label{fig-06}
\end{figure}

\textit{Co-based Heusler compounds.} There is some degree of scatter
in the AMR values reported for Co-based Heusler alloys. For example
the values for Co$_2$MnGa were found to lie~\cite{Sato:2019} between
-2.5 \% and +0.75 \% depending on current direction and precise stoichiometry.
The former dependence can be analysed in terms of crystalline and
non-crystalline terms (see discussion later in this Section) as in
Fig.~\ref{fig-09} but a meaningful comparison between these two, i.e.
epitaxial~\cite{Sato:2019} and sputtered~\cite{Ritzinger:2021} samples,
requires also the knowledge of background resistivity~\cite{Sato:2018}
proportional to $R_0$ from~(\ref{eq-01}).

On the other hand, once the current direction is fixed (here, along [110]
crystallographic direction) we often arrive at similar characteristics of AMR
even for different compounds: measurements of Co$_2$MnGa by
Ritzinger et al. \cite{Ritzinger:2021} and of Co$_2$FeAl by
Althammer \cite{Althammer} show negative AMR which decreases with
temperature and is quite small ($\approx 0.1 - 0.2 \%$). Several other
examples are given in Fig.~\ref{fig-06}.

\begin{figure}[h]
  \centering\includegraphics[scale=0.4,angle=0]{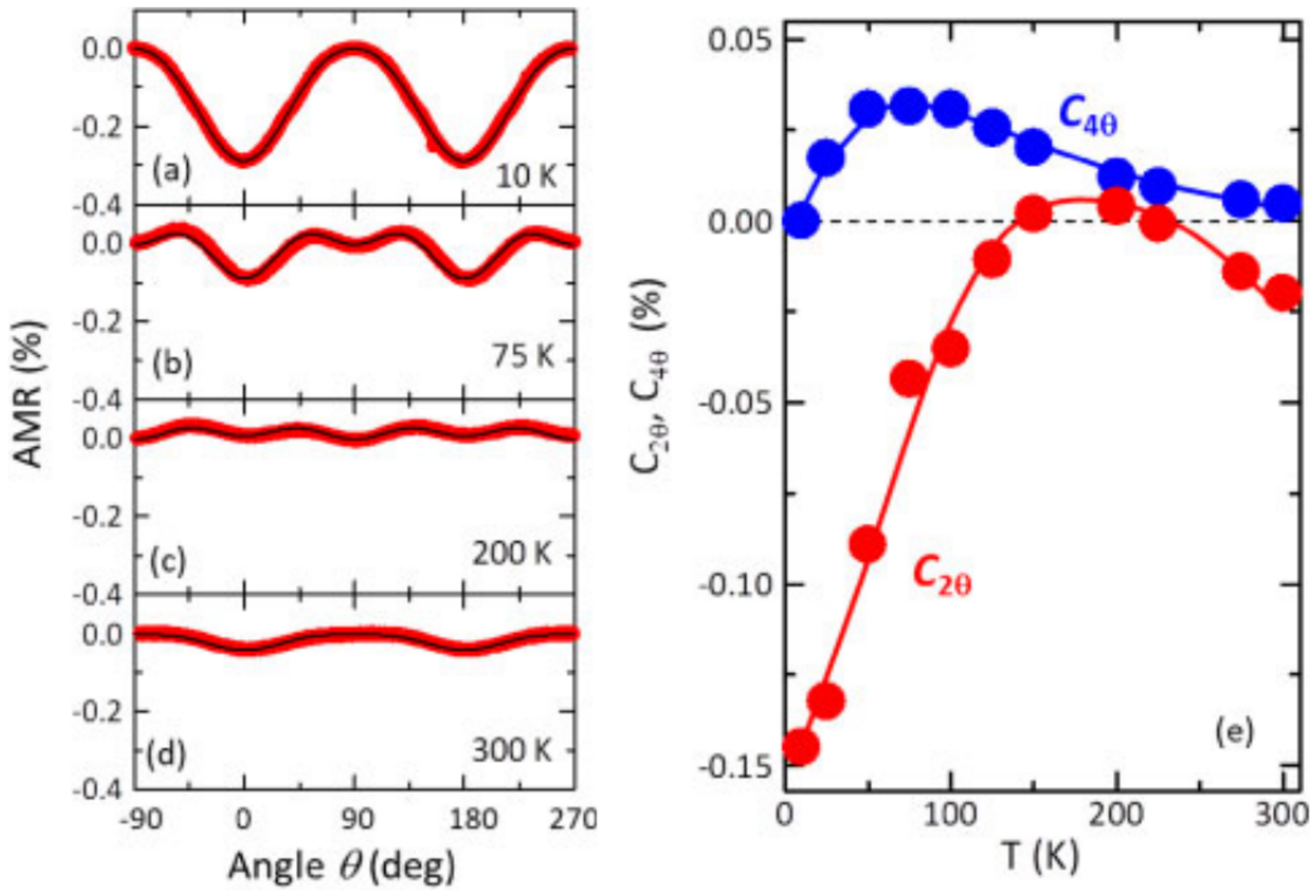}
\caption{\lic{NO (Applied)}{Oogane}{f05l}  
    AMR in Co$_2$MnSi. (a)-(d) AMR for different temperatures between 10 and 300 K, (e) temperature evolution of the two- and the four-fold fourier-component of the AMR. Reproduced from Ref.~\cite{Oogane:2018}.}
  \label{fig-05}
\end{figure}

\begin{figure}[h]
\centering\includegraphics[scale=0.8,angle=0]{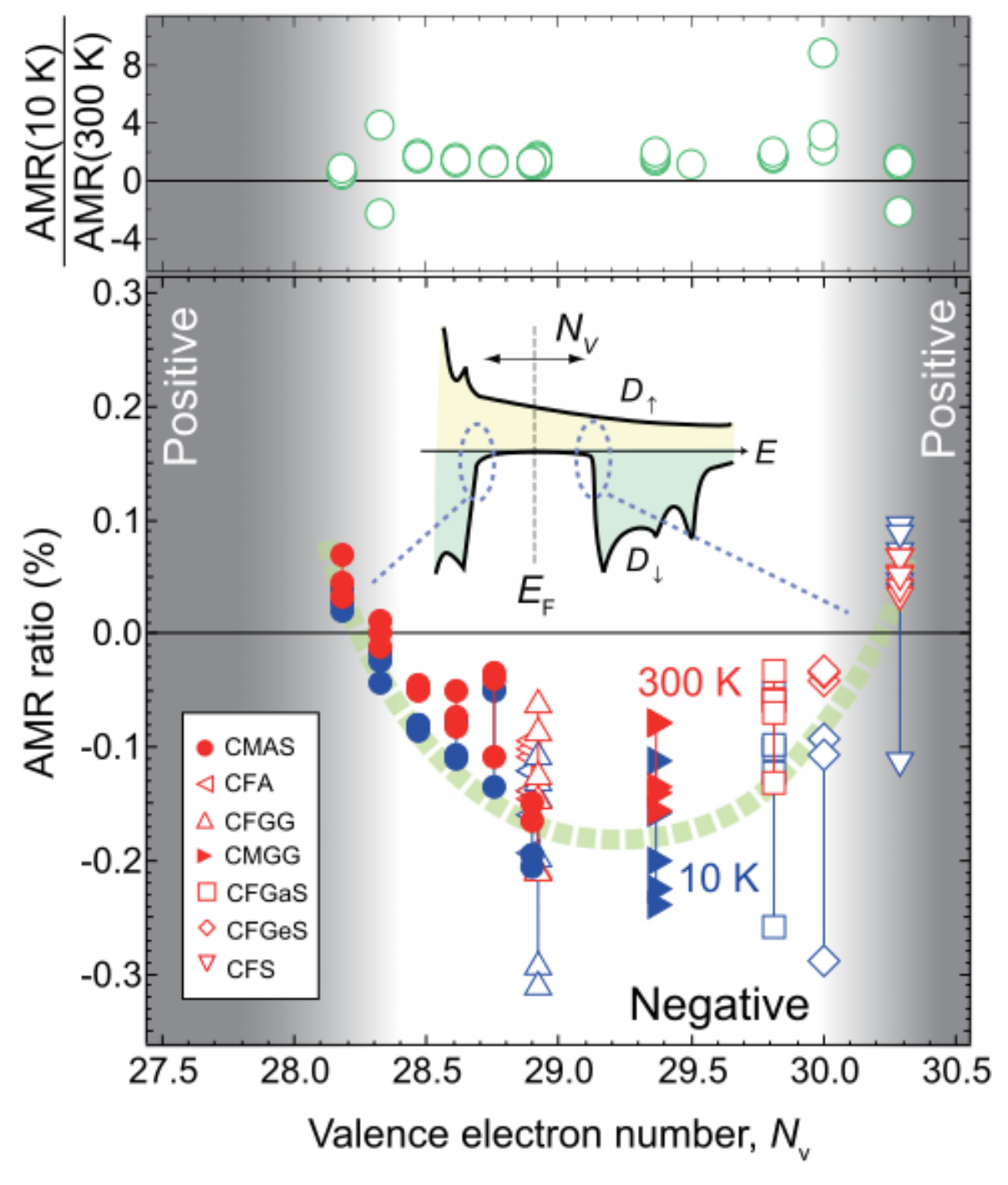}
\caption{\lic{YES}{Sakuraba}{fDOS}
		Valence electron number $N_V$ dependence of AMR ratio in all
                Co$_2$MnZ and Co$_2$FeZ films??. The inset shows the respective density
                of states. The upper part shows the ratio of AMR ratios at 10 K to 300 K. Reproduced from Ref.~\cite{Sakuraba:2014}.}
	\label{fig-14}
\end{figure}

Investigation of AMR in Co$_2$FeZ and Co$_2$MnZ with Z = (Al, Si, Ge,
Ga) and current along [110] found both negative and positive AMR,
depending on the total number of valence electrons $N_v$. If that
number was between 28.2 and 30.3, a negative AMR was reported,
otherwise a positive. According to bandstructure calculations, in
between $N_V$ of 28.2 and 30.3, it corresponds to half-metallicity
\cite{Sakuraba:2014} as can be seen in Fig.~\ref{fig-14}. The reported
AMR ratios in this paper are relatively small as in the other
papers. An equivalent result was achieved in Co$_2$Fe$_x$Mn$_{1-x}$Si:
Here the AMR is negative for x $\leq$ 0.6 and positive for x $\geq$
0.8, which is explained by a transition from minority conduction to
majority conduction and thus interpreted as a possible sign for
half-metallicity as well \cite{Yang:2012}. Similarily in Co$_2$FeSi,
the AMR ratio was determined for different samples distinguished by
their annealing temperature: Above 600$^\circ$ C the AMR is negative,
up to 600$^\circ$ C it is positive with the same explanation as before
\cite{Yang:2013}. The AMR ratio in Co$_2$(Fe???Mn)Si,
Co$_2$(Fe???Mn)(Al???Si) and Co$_2$(Fe???Mn)Al was reported to be $\approx
-0.2 \%$ for low and room temperature \cite{Yako:2015}.

While the majority of studies focuses on AMR ratio and sign, the symmetry of these compounds are also a puzzling topic: In Co$_2$MnGa \cite{Ritzinger:2021}, Co$_2$FeAl \cite{Althammer} and Co$_2$MnSi \cite{Oogane:2018}, the AMR showed a complex signal comprising of non-crystalline and crystalline terms. However, the division into non-crystalline and crystalline is usually not made and the AMR is only described in terms of $\cos(4\phi)$ and $\cos(2\phi)$ contributions (be $\phi$ some angle of rotation). The appearing symmetries makes AMR in these materials much more complex as e.g. in simple transition metals, where normally only two-fold symmetries are found. An example of such a rather complex signal in Heusler compounds can be found in Fig. \ref{fig-05}.

The 4-fold contributions in these signals are too strong to be
ascribed to MCA solely. In a theoretical study by Kokado and Tsunoda
\cite{Kokado:2015}, it was suggested that a tetragonal distortion of
the crystal structure can introduce such a 4-fold crystalline AMR
contribution. Please note, that the Heusler alloys per se have cubic
crystal structure, but in thin-films the substrate usually introduces
a small tetragonal distortion. Still, despite this explanation being
plausible, the complex temperature dependence of the 2- and 4-fold
contributions \cite{Althammer,Ritzinger:2021} asks for further investigation.

In conclusion, a couple of observations can be made for AMR in
Co-based Heusler alloys: Firstly, the AMR ratio is generally
decreasing for increasing temperature. Secondly, the AMR is very
small, often well below 1 \% and only in some specific configurations
(low temperature, favorable stroichiometry) it reaches up to  $\approx
2\%$. Thirdly, the AMR ratio given by~(\ref{eq-01}) is usually
negative, however can be tuned to be positive. This behavior can be
seen consistently in various studies and appears to be a general
property of this material class of the Co-based Heusler
alloys. Various "phenomenological" explanations for the sign change of
AMR are given, e.g. a dependence on annealing temperature, Fe-content,
Co-content, current direction and $N_v$. These explanations are rather
diverging and not allowing for a consistent conclusion. 
On a microscopic level, however, the various studies can be summarized
quite well: as long as the compounds are having a half-metallic
character / showing minority conduction, the AMR is negative.
In the case of majority conduction and metals not fully polarised on
the Fermi level, the AMR becomes positive. It appears to be the case, that the Co-based Heuslers investigated here are all \textit{by default} (= in an ideal configuration) half-metallic, but can be all tuned to lose this half-metallic character (this tunig was done by considering the phenomenlogical aspects, such as annealing temperature). The theoretical model used to explain it was developed by Kokado and Tsunoda~\cite{Kokado:2012}. \\


\textit{Non-Co-based Heusler and semi-Heusler compounds.}
Just as with transition metal
alloys, Heusler materials~\cite{Felser-book} span a vast range of compounds:
magnetic Heusler alloys include NiMnSb~\cite{Ciccarelli:2016_a} or
Ru$_2$Mn$_{1-x}$Fe$_x$Ge, which is a ferromagnet for x = 1 (no Mn) and an antiferromagnet for x = 0 (no Fe). For x = 0.5 an anisotropy in the MR is observed with a MR of -4\% and +2\% under parallel and perpendicular configurations of applied field and applied current, respectively. It was speculated, that this (Anisotropic) MR might stem from random alignment of ferromagnetic domains. For x = 0 and x = 1 no MR was found~\cite{Mizusaki:2009_a}.

\subsection{Two-dimensional electron gases}
\label{sec_2DEG}

\textit{Introduction.} A two-dimensional electron gas (2DEG) can form
on various interfaces: 
surface of liquid helium, classical semiconductor heterostructures or
certain transition metal oxid interfaces (TMOI). The textbook example
of such a TMOI is a SrTiO$_3$/LaAlO$_3$ (STO/LAO) interface, where the
two perovskites individually are non-magnetic insulators \cite{Shalom_1}. The
research interest in TMOIs can be broadly speaking divided into three
categories: i) General understanding of the electronic structure,
magnetism and related effects, ii) understanding of the
superconductivity \cite{Shalom:2009, Shalom_2, Huijben:2009}
(transition temperature is typically 350 mK \cite{Lebedev:2020}) and
iii) possible development of applications, such as quantum-matter
heterostructures \cite{Boschker:2017_a}.
Regarding AMR in TMOI-hosted 2DEGs, it is important to distinguish whether
the transport anisotropy occurs due to orbital effects~\cite{Bovenzi:2017}
as discussed in Sec. 1(d) on general level, or if it is indeed related to
magnetism. Hysteretic magnetisation loops observed in STO/LAO structures
grown at suitable oxygen pressure~\cite{Ariando:2011} can be taken as
a hint of the latter yet the magnetoresistances shown in Fig. 3 of that
reference clearly show that even here, the orbital effects are strong. 
On the other hand, longitudinal and transversal magnetoresistance (MR)
showing similar behaviour of LTO/STO (LTO = LaTiO$_3$) at stronger
magnetic fields (compare Figs. 2d~and~e in Ref.~\cite{Lebedev:2020})
can be taken as an argument that the latter are {\em not} dominant.
The focus of many publications lies on LAO/STO interfaces, whose
results are discussed in the following. A summary of AMR in other
TMOI-hosted 2DEGs (including LTO/STO~\cite{Lebedev:2020})
can be found at the end of this section. 

\begin{figure}[h]
  \includegraphics[angle=0,scale=0.8]{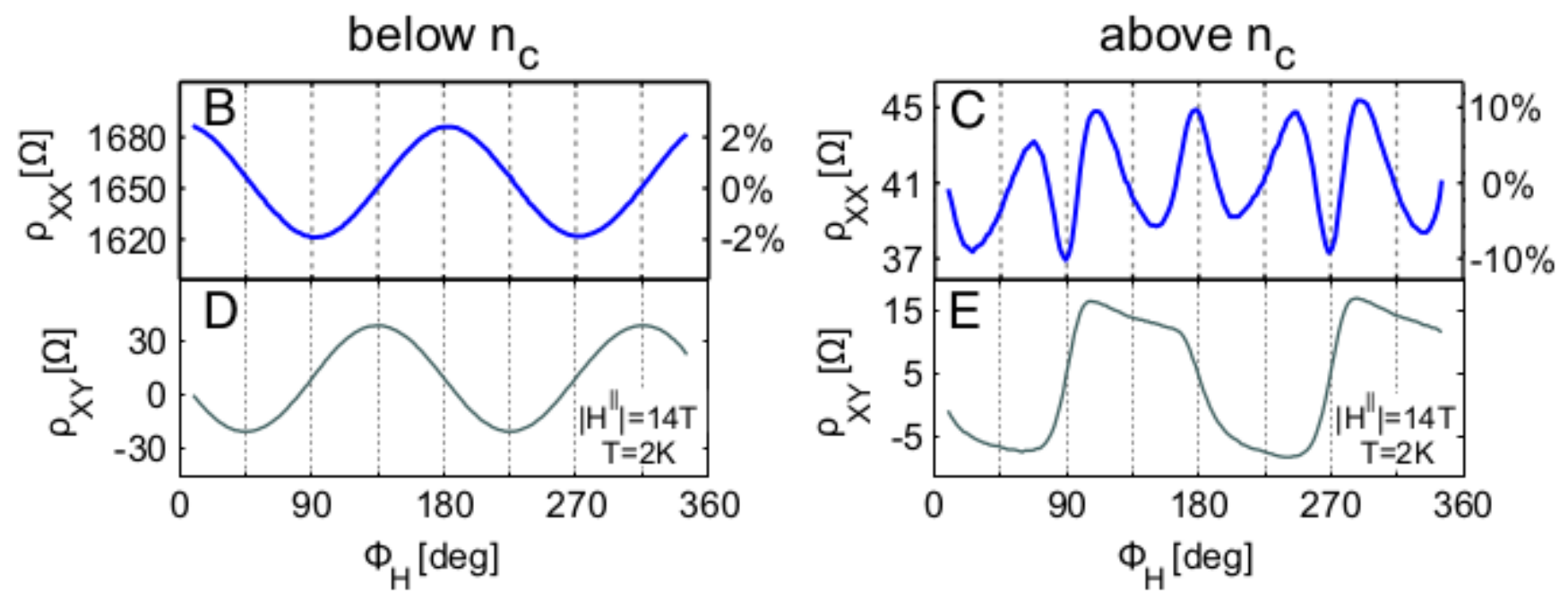}
\caption{\lic{YES}{PNAS}{f08l} 
    (B,D) AMR in a 2DEG on LAO/STO interface for electron density below (B) and above (D) the critical value.
    Reproduced from Ref.~\cite{Joshua:2013}.}
  \label{fig-08}
\end{figure}

\begin{figure}[!h]
	\centering\includegraphics[scale=0.6,angle=0]{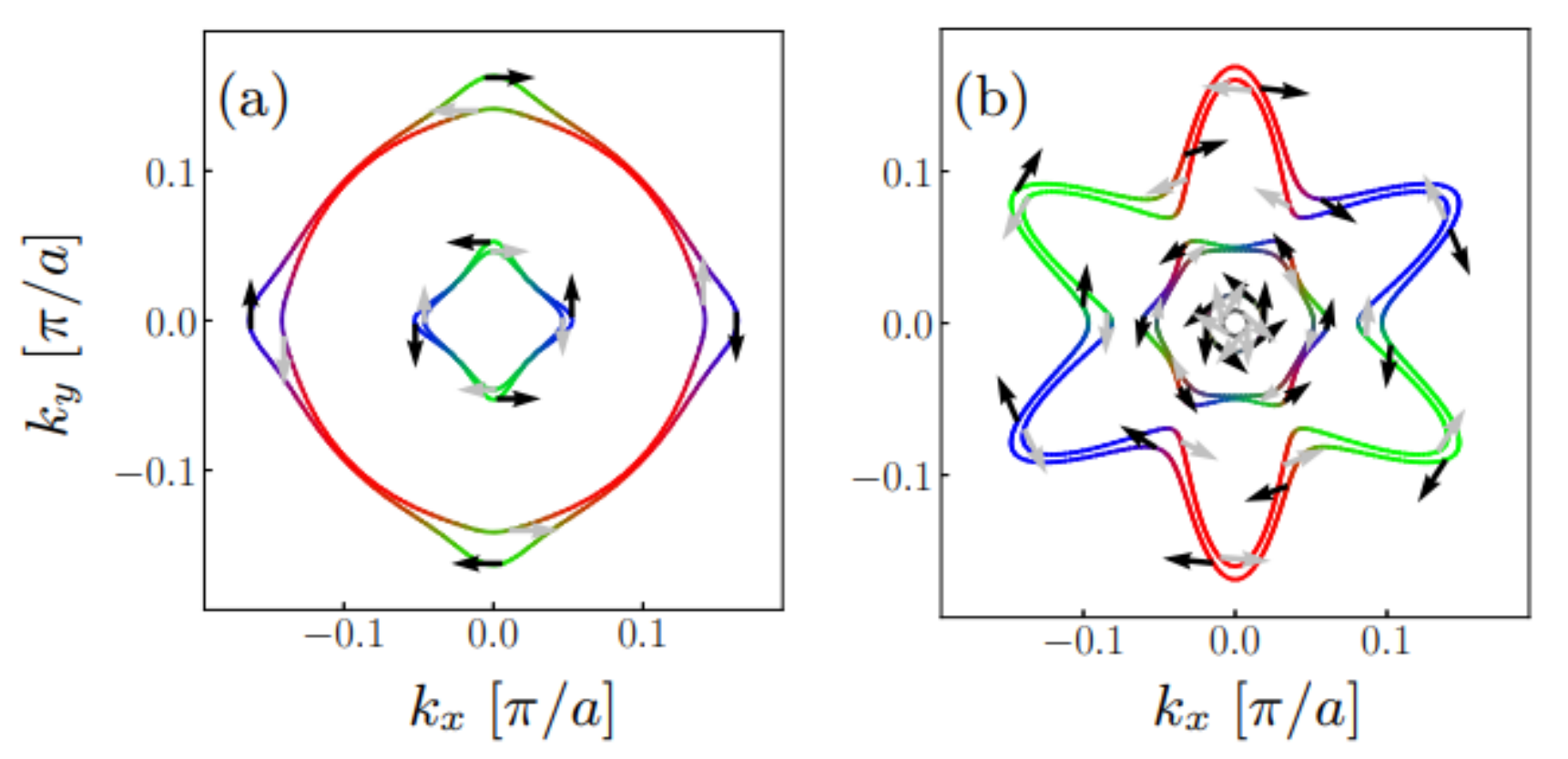}
\caption{\lic{YES}{Boudjada}{f13l}
Fermi surfaces for low-(electron) density spin-orbit coupled
2DEGs at zero magnetic field, with colors indicating orbital content
(yz-blue, zx-green, xy-red), and Rashba spin texture indicated by
black/gray arrows for opposite chiralities. (a) (001) 2DEG and (b) (111) 2DEG.
Both Fermi surfaces are highly anisotropic. Reproduced from Ref.~\cite{Boudjada:2019}}
	\label{fig-13}
\end{figure}

\textit{AMR in LAO/STO.} On qualitative level, the AMR of a 2DEG
at the LAO/STO interface can exhibit two types of behaviour, see
Fig.~\ref{fig-08}. This was attributed to a
phase transition when going to low temperatures \textit{T} and high carrier densities \textit{n}. A positive and two-fold AMR was found for temperatures $T > 35$ K, while for lower $T$ and higher $n$ negative and showing higher orders up to 6-fold symmetry in (111) and (110) interfaces~\cite{Rout:2017, Joshua:2013, Miao:2016, Shalom:2009}
was found.\\
The absolute value of the AMR ratio is not the major point of discussion and should be taken with maximum caution, due to its huge dependence on temperature \cite{HarsanMa:2017}, current density (AMR increasing with increasing \textit{n} \cite{Rout:2017}), \textit{B}-field strength \cite{HarsanMa:2017, Miao:2016} and many open points in the understanding of the inner workings of AMR in these materials. The AMR was reported to be larger in the low-\textit{T} high-\textit{n} phase ($\approx 2\%$ below and $\approx 10\%$ above the critical $n$)~\cite{Joshua:2013}. A large value of 110 \% was reported for some $[1\bar{1} 0]$ oriented samples grown under low oxygen pressure with B = 9 T~\cite{HarsanMa:2017},
which was understood in terms of oxygen vacancies leading to stronger
orbital polarizations and producing a more anisotropic Fermi surface (FS)
which leads to larger AMR \cite{HarsanMa:2017}. Also, the
band structure and thus the FS and the AMR are strongly dependent on the sample orientation~\cite{HarsanMa:2017} and oxygen pressure during growth.\\
%
In calculations, the AMR is frequently linked to a strong anisotropy of the FSs as exemplified in Fig.~\ref{fig-13}~\cite{Joshua:2013, Bovenzi:2017, Boudjada:2019, HarsanMa:2017}. Although this means that the AMR is intrinsic, the distinction between intrinsic and extrinsic AMR in these studies is usually not made. The harmonics of the AMR, i.e. the strength of the 2-, 4- and 6-fold are not directly linked to the symmetry of the FS~\cite{Boudjada:2019}.\\
The electronic structure at the FS is different between the low-\textit{n} and high-\textit{n} regimes \cite{Bovenzi:2017,Boudjada:2019} and is also sensitive to the crystallographic direction of interface \cite{Boudjada:2019, Miao:2016}. The anisotropy appears to be driven by interband scattering, which is surpressed in the low-\textit{n} regime \cite{Boudjada:2019}. The $t_{2g}$-orbitals and broken inversion symmetry are generally a central part in the modelling of LAO/STO interfaces \cite{HarsanMa:2017, Miao:2016, Joshua:2013, Bovenzi:2017,Boudjada:2019}.

%

\textit{Other Materials.} Apart from the much investigated LAO/STO
interface, 2DEGs at a TMOI can be found in other material combinations,
for example
LVO/KVO (LaVO$_3$/KTaO$_3$)~\cite{Wadehra:2020}, LVO/STO~\cite{Tomar:2021},
LTO/STO~\cite{Lebedev:2020} and CZO/STO~\cite{Chen:2015_a} interfaces
where CZO stands for CaZrO$_3$ (the last mentioned system stands out by being
non-polar without strain). AMR was only studied in the first two examples;
sometimes, anisotropic data is shown~\cite{Lebedev:2020} (longitudinal
and transversal MR as mentioned above), although not being referred to
as AMR in the respective publication. The results are shortly
summarized in the following. Please note that the similarity of the
2DEGs in LTO/STO and CZO/STO do still suggest the existence of similar
AMR phenomena, which yet have to be investigated.

In low temperature measurements in (001)-interfaces of LVO/KVO \cite{Wadehra:2020} and LVO/STO \cite{Tomar:2021} a low-field two-fold AMR turned into a high-field four-fold AMR. In case of a (111)-interface of LVO/STO the high-field AMR was six-fold. AMR in LVO/STO showed a strong field- and temperature-dependence. The larger four-fold AMR persisted up to 150 K while the six-fold AMR persisted up to 20 K, similar to the situtation in LAO/STO \cite{Tomar:2021}. While no profound explanation was given for the LVO/KVO interface \cite{Wadehra:2020}, it was suggested that AMR in LVO/KVO is due to an anisotropic FS, similar to the situation in LAO/STO \cite{Tomar:2021}.

\subsection{...and all the rest}
\label{sec_MatRest}

The previous five sections of this chapter were devoted to the material classes showing the most important and remarkable results in the field of AMR. This not nearly a complete picture of the universe of AMR. In the following section we are going to discuss briefly results in several other material classes. \\

\textit{Fe-based alloys}: Apart from iron-cobalt and iron-nickel
alloys which were discussed already in Sec.~3(a),
Berger et al.~\cite{Berger:1988_a} investigated AMR also in
Fe-Cr and Fe-V and split the AMR contributions into parts due to phonon
and impurity scattering (see also discussion of Ref.~\cite{Berger:1988_a}
below in the context of CoPd alloy). It was suggested
that in alloys with strong scattering the AMR changes sign when the
impurity scattering is maximal. According to this study, a change of
3d-DOS does not account for all of the observed behavior.

In Fe$_{0.8}$Ga$_{0.2}$ it was found that the AMR is two-fold and for in-plane (out-of-plane) configuration at a magnetic field of 500 mT (8 T) showed negative (positive) AMR. Interestingly, with increasing temperature the AMR is constant (decreasing). The AMR ratio is slightly larger than 0.1 \% (between ca. 0.2 and 0.5 \%). The perfect two-fold shaped AMR curves were interpreted as a sign that saturation magnetization was reached \cite{Granada:2016_a}. \\
Properties of NiFeCr alloys such as AMR ratio, low-temperature resistivity and $T_C$ depending on the Cr concentration are listed in Table 1 of Ref ~\cite{Chakraborty:1998_a}. A maximum AMR of $0.76\%$ is found in the sample with the lowest Cr concentration of $2\%$. Increasing the Chromium content leads to a rapid decrease of AMR ratios until the AMR almost vanishes for concentrations higher than $18\%$. Please note, that this study is using the term \textit{ferromagnetic anisotropy of resistivity (FAR)} instead of AMR. The rapid decrease of the AMR ratio is accompanied by a drop of $T_C$, from 778 K for $2\%$ concentration to 48 K at $21\%$ concentration~\cite{Chakraborty:1998_a}.\\

\textit{Other alloys or structures involving transition metals.} 
The AMR of Co-Pd alloys was investigated in Ref.~\cite{Jen:1992_a} for
various cobalt concentrations $x$ and its temperature dependence was
analysed in terms of Parker plots (as discussed in the introduction of
Ref.~\cite{Berger:1988_a}). A maximum ratio of almost $8\%$ was reported at low temperatures for almost equal concentrations of Co and Pd. The results were interpreted with the framework of s-s- and s-d-scattering, splitting the resistivity into contributions of spin up- and down, s-s and s-d-scattering and phonon- and impurity-contribution to the AMR~\cite{Jen:1992_a}. It can be seen as an advancement of the theory of Campbell, Fert and Jaoul discussed in Sec. 2\ref{sec_TM}.
\\
Calculated values of the AMR ratio and the residual resistivity of Co-Pd and Co-Pt alloys as a funcitomn of the Cobalt-concentration is shown in Figs. 5 and 3, respectively, of Ref.~\cite{Ebert:1996_a}. The values are compared to experimental values from various studies, which showed the accuracy of the calculation. In case of Co-Pt the AMR reaches values of up to $1\%$, while in the Co-Pd case the AMR shows a maximum of $6\%$ (calculation) or $8\%$ (experiment). The AMR is starkly decreasing for very low Co-content~\cite{Ebert:1996_a}.
Note that even for concentrations as low as 3\% of cobalt, palladium alloys 
remain ferromagnetic~\cite{Stampe:1995_a} and the AMR can
  be reasonably modelled assuming $|J|=43$~meV for the coupling between
  magnetic moments and highly conductive $s$-electrons.
  
The in-plane and out-of-plane AMR of Nickel sandwiched by Platinum was experimentally investigated and the symmetry of the AMR discussed~\cite{Philippi-Kobs:2019}. The nickel films are fcc textures with a (111) surface and have a thickness between 2 and 50 nm, while the Platinum layers 5 and 3 nm thick. The in-plane AMR shows only two-fold symmetry as expected for an isotropic polycrystalline sample. The out-of-plane AMR shows pronounced four-fold- and six-fold-symmetries for nickel thickness $\geq 6$ nm. The h.o. symmetries were explained using phenomenlogical symmetry-based arguments \cite{Doring:1938} due to (111) textured interface and Fuchs-Sondheimer theory for scattering at interfaces. All results were obtained at room temperature~\cite{Philippi-Kobs:2019}.\\
The symmetry of Fe-monolayers on a GaAs interface changed depending on the number of monolayers. While for 8 monolayers, a four-fold component was dominant, with decreasing number of monolayers to 6 and 4 the four-fold component decreased. This was attributed to a change of symmetry due to transitioning from bulk-like to interface-like symmetry~\cite{Hupfauer:2015}.\\

\textit{The perovskite Iron Nitride Fe$_4$N and the derived materials} CuFe$_3$N \cite{Shi:2020} and Mn$_4$N \cite{Kabara:2017}. For the iron
nitride case, we can distinguish into in-plane AMR
\cite{Tsunoda:2009,Tsunoda:2010, Kabara:2014} and
\textit{transverse AMR} (Magnetic field $\vec{H}$ rotated in the plane perpendicular to the current $\vec{j}$)
\cite{Kabara:2016}.
FeN$_4$ in the matrix of Fe-doped GaN also exhibits AMR~\cite{ANQ2019}. \\
In all samples a four-fold component of the AMR was found, for example in in-plane Fe$_4$N below 30 K \cite{Kabara:2014}, and it is almost vanishing at higher temperatures. In \textit{transverse} AMR of Fe$_4$N and in Mn$_4$N the four-fold-component is dominant for low temperatures.\\
All samples show negative AMR at low temperatures. The AMR in Fe$_4$N (in-plane) and in CuFe$_3$N remain negative, while Fe$_4$N (\textit{transverse}) and Mn$_4$N showing positive AMR for temperatures above approx. 50 and 100 K, respectively.
Low temperature AMR ratios for Fe$_4$N and Mn$_4$N scatter between approximately $-0.75 \%$ \cite{Kabara:2014} and $-7 \%$ \cite{Tsunoda:2010} in iron nitride and around 2 \% in manganese nitride \cite{Kabara:2017}). While AMR ratios scatter in general, an increasing AMR ratio for increasing annealing temperature was reported in iron nitride \cite{Kabara:2014}. In the ferromagnetic anti-perovskit $\gamma'-$CuFe$_3$N, low temperature values in the range of $-0.067$ to $-0.336 \%$ were reported, at higher temperatures dropping to $0.003 \%$. \\
In iron and manganese nitride the decrease of the AMR coefficients with increasing temperature show a kink at about 50 K, changing from rapid to moderate decrease. No explanation was given. \\
The results were discussed in the framework of sd-models \cite{Kokado:2012}. Negative (positive) AMR ratios were linked to minority (majority) spin conduction, while appearing four-fold-symmetries were linked to possible tetragonal distortion. In Fe$_4$ this was suggested to be due to anisotropic thermal compression \cite{Kabara:2016}.\\

\textit{Some more perovskites.} Metallic SrRuO$_3$ exhibits negative
magnetoresistance
as expected for ferromagnets, and
its form for parallel and perpendicular configuration of magnetisation and
current confirms this is AMR rather than an orbital effect~\cite{Klein:1998_a},
the former being negative and achieving quite large values of $\approx 25\%$
at low temperatures. The AMR is decreasing slowly for low temperature and steeply for higher temperatures above $\approx 100$~K as it is approaching and surpassing the Curie temperature of $\approx 140$~K and it does not show any enhancement in the vicinity of the Curie temperature, which is contrary to the results in manganites and was attributed to a missing of the Jahn-Teller effect~\cite{Herranz:2004}. 
The magnetoresistance depends sensitively on strain, however~\cite{Rao:1998_a}.
AMR and PHE were compared at low temperatures and it was found that the AMR is almost double as large as the PHE with $\approx 14\%$ and $\approx 7\%$, respectively~\cite{Haham:2013},
implying sizable crystalline AMR terms.\\
The nonmagnetic SrIrO$_3$ (SIO) shows AMR for temperatures below 20 K which was interpreted as a sign of a possible ferromagnetic ordering emerging at low temperatures induced by local structure distortion due to lattice strain. The presented data was close to a two-fold AMR. More precisely, the fitting process yielded AMR $\propto \cos(1.75 \phi)$ \cite{Chaurasia:2021}. \\
Interestingly, bulk SrIO$_3$ does not show such behavior and the here investigated film is a thin-film on a SrTiO$_3$ substrate \cite{Chaurasia:2021}. In the previous section, we discussed various examples of thin-films on STO substrats forming a 2DEG at about the same temperature accounting for the transport effects. In our judgement, this could possibly account for the emergent AMR at low temperatures.\\


\textit{Manganites} form a large class of perovskite materials ranging
from (the more common) antiferromagnets (such as CaMnO$_3$~\cite{Zeng:1999_a})
to ferromagnets (less common in ternary~\cite{andGoodenough21} and
well-established in numerous quaternary systems described below).
Often~\cite{Herranz:2004}, this material class is defined as compounds
of the form X$_a$Y$_b$MnO$_3$, where X and Y are a trivalent and divalent cation, respectively, with their respective concentrations $a$ and $b$. The main part of manganites discussed here are based on Lanthanum, for which the second element $Y$ is either Ca \cite{Bibes:2005, Li:2010, ODonnell:2000, Yang:2021_a, Xie:2013_a, Sharma:2014}, Pr \cite{Kandpal:2016, Alagoz:2014}, Sr \cite{Wong:2014} or Ag \cite{Infante:2006}). Among the other materials are Nd$_{0.51}$Sr$_{0.49}$MnO$_3$ \cite{Kumar:2011} and Sm$_{0.5}$Ca$_{0.5}$MnO$_3$ \cite{Chen:2009_a}. 
In many of the studies a STO substrate was used \cite{Alagoz:2014, Chen:2009_a, Li:2010, Bibes:2005, ODonnell:2000, Xie:2013_a}, while sometimes also other substrates such as LAO \cite{Alagoz:2014} and BaTiO$_3$ (BTO) \cite{Xie:2013_a} were reported. The role of the substrate in the results is here solely attributed to the strain it applies on the manganite layer. 
Lathanum-based oxides ordering in a perovskite structure on a STO substrate resemble on the first glance the LAO/STO samples discussed in the previous section in terms of the 2DEGs. The difference is that LAO is a non-magnetic isolator, where magnetism and transport are only occuring at the interface with the substrate.
\\
The AMR is usually reported to be two-fold, however also four-fold AMR was reported. While the four-fold symmetry was reported to be robust in La$_{2/3}$Ca$_{1/3}$MnO$_3$~\cite{Li:2010}, it only appeared on a tensile strained La$_{0.4}$Sr$_{0.6}$MnO$_3$ sample on a STO substrate~\cite{Wong:2014}. Other substrates showed two-fold AMR for the same material. 
Arguably the most attention was paid to ferromagnetic LCMO (with X=La, Y=Ca 
and $a,b$ in $2:1$ ratio) where the colossal magnetoresistance (CMR) occurs,
and here, the AMR at low temperatures~\cite{Xie:2013_a} is clearly observable
but small. At higher temperatures a peak was found in slightly
off-stoichiometric LCMO, La$_{0.7}$Ca$_{0.3}$MnO$_3$ films~\cite{ODonnell:2000}. In the former case \cite{Xie:2013_a} it was ascribed to be due to strain from the BTO substrate.\\
The sign of the AMR is in most cases predominantly negative. However, there are studies which report a sign change of the AMR as a function of temperature \cite{Chen:2009_a, Yang:2021_a, Alagoz:2014, Infante:2006} and others which report exclusively negative results \cite{Bibes:2005, Kandpal:2016}. In the study of Xie \textit{et al}. \cite{Xie:2013_a}, a 80 nm thick-sample on a BTO substrate was reported to show a sign change, while the other samples are solely negative. The sign change of the AMR was sometimes linked to a change of the easy axis with temperature.\\

\begin{figure}[h]
\centering\includegraphics[scale=0.5,angle=0]{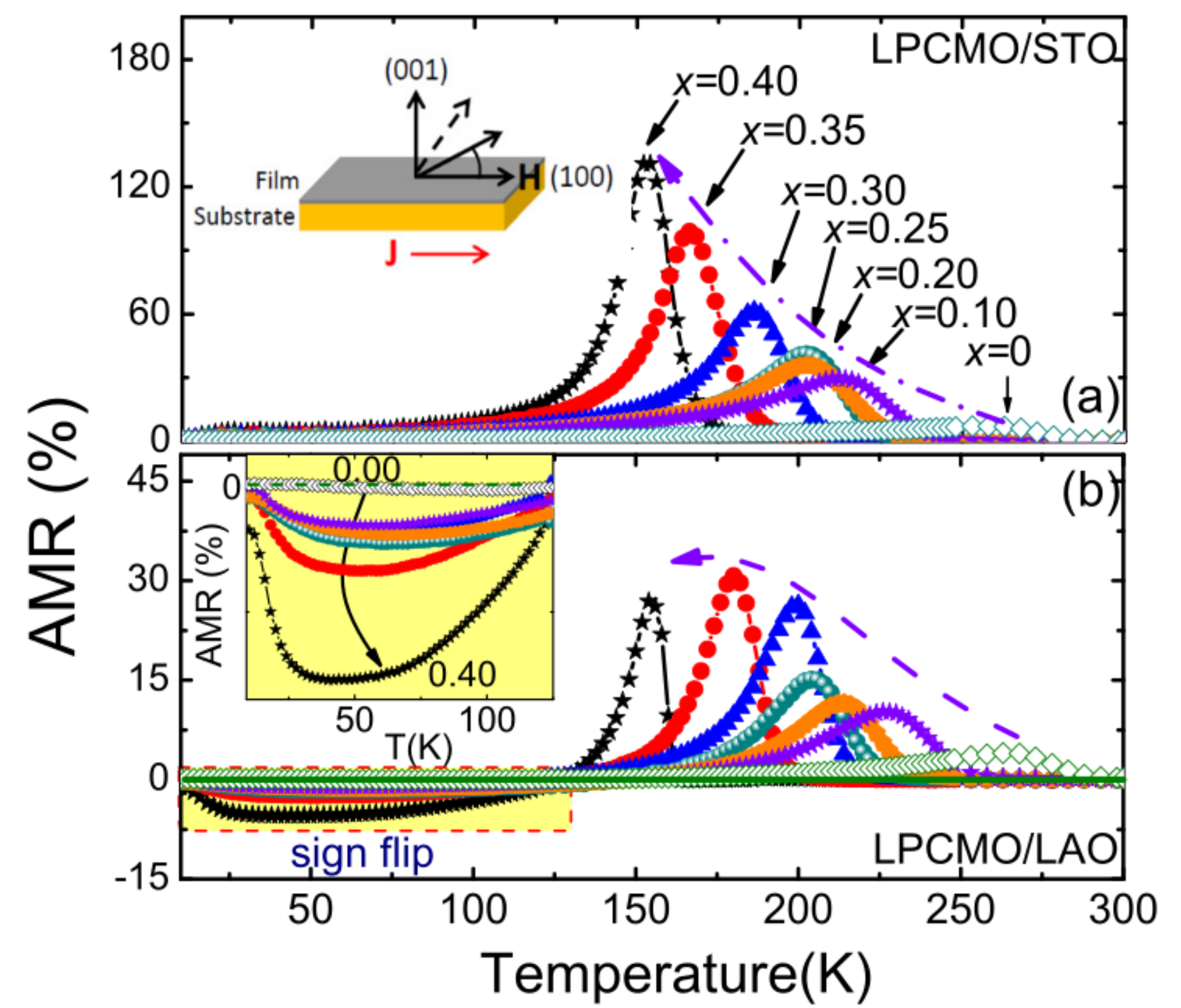}
\caption{\lic{YES}{Alagoz}{f15l}
AMR in La$_{0.7-x}$Pr$_x$Ca$_{0.3}$MnO$_3$.
Temperature dependence of AMR measured in a field of 1.1 T at
doping levels x = 0, 0.10, 0.20, 0.25, 0.30, 0.35, and 0.40 for films grown on
(a) SrTiO$_3$ and (b) LaAlO$_3$ substrates. Dashed curves in both figures show
the expected dependence of the AMR$_{max}$ on doping. The change of sign of
the AMR with an increasing doping is shown in the rectangular yellow area.
Insets in (a) and (b) show the direction of magnetic field H and the direction
of the current J, and the expanded view of the AMR at low temperatures for
LPCMO/LAO doped films, respectively. The figure is reproduced from Ref.~\cite{Alagoz:2014}}
	\label{fig-15}
\end{figure}

Magnitudes of the AMR ratio are scattered between approx. $0.1 \%$
in LCMO~\cite{Bibes:2005} and 'colossal' values in excess of $100 \%$
for La$_{0.3}$Pr$_{0.4}$Ca$_{0.3}$MnO$_3$ at its peak value approx. below 150 K \cite{Alagoz:2014} on a STO substrate as can be seen in Fig.~\ref{fig-15}(a). AMR values were reported highly sensitive to e.g. sample composition, type of substrates \cite{Alagoz:2014} - and thus strain - and current directions \cite{Bibes:2005}. An example of the dependence of AMR on the substrate and the doping levels is shown in Fig.~\ref{fig-15}.

In an absolute majority of the studies, the low-temperature AMR
increased with increasing temperature in clear contrast to the usual
behavior and peaked just below the metal-insulator transition
temperature. Above this temperature, the AMR ratio decreased
rapidly. Microscopic mechanism of AMR in these materials is clearly
distinct~\cite{Egilmez:2008} from that in conventional ferromagnets
(such as alloys of transition metals discussed previously).
These temperatures are scattering from low temperatures \cite{Kandpal:2016} up to almost room temperature~\cite{Yang:2021_a}, however usually somewhat lower than the latter.
Exceptions to the high temperature peak are rare and occur for example in thin-film samples of LCMO on a BTO substrate \cite{Xie:2013_a}, in La$_{0.4}$Sr$_{0.6}$MnO$_3$ \cite{Wong:2014} and in a polycrystalline Nd$_{0.51}$Sr$_{0.49}$MnO$_3$ sample \cite{Kumar:2011}, while the single-crystalline samples of the latter show the characteristic peak.\\
The explanation for the characteristic behavior is usually linked to strain resulting in orbital deformation via the Jahn-Teller effect \cite{Chen:2009_a, Wong:2014, ODonnell:2000, Alagoz:2014, Infante:2006}. While other authors offered explanations linked to double exchange \cite{Xie:2013_a} or to a magnetic liquid behavior \cite{Kandpal:2016}.

It is worth noting that the 'colossal' peak values of AMR can be usually observed in the vicinity of the MIT-temperature, where also the CMR effect occurs. Thus, explanations of these large AMR values have to be taken with some caution as neither the CMR effect is fully understood itself. In the latter case mainly due to a lack of quantitative theory describing the MIT and the subsequent insulating phase, which in our judgement might be problematic in the description of the AMR as well.

\textit{Other conductive oxides.} 
In Fe$_3$O$_4$ (magnetite), AMR was used to refute predicted
half-metallicity~\cite{Ziese:2000_a} and later,
a strong crystalline contribution
was reported \cite{Li:2010_b}. The origin of two- and four-fold AMR in
Magnetite was linked to magnetic anisotropy and to scattering far away
and near the antiphase boundaries, respectively. This oxide can be
alloyed with nickel while still remaining conductive: AMR in
Ni$_{0.3}$Fe$_{2.7}$O$_4$ shows a strong four-fold component as
well~\cite{Li:2010_b}.
Another example of a conducting oxide is the AFM RuO$_2$. Here, angle-dependent spin-torque ferromagnetic resonance measurements (ST-FMR) in Fig. 3b and d of \cite{Bai:2022_a} resemble AMR and indicate the existence of the effect in the material.\\

\section{Applications and Further Topics}
\label{ch_Appl}

There is a broad range of opportunities to exploit the AMR and also,
to go beyond magnetisation-controlled DC resistance. In the following
we will shortly discuss both industrial and scientific applications of this effect,
ranging from the well-known hard-disk read heads in pre-GMR era to
subtle techniques for detection of spin relaxation in ferromagnets. Related
phenomena in optics and thermoelectricity will be mentioned.



\subsection{Scientific applications}
\label{sec_ApplSci}

Since AMR gives the dependence of the resistivity on the magnetization direction, it can be used as a means of \textit{magnetometry}. This is the main application of AMR in a scientific context and a few examples are given in the following paragraph. The second paragraph of this section summarizes a variety of other applications of AMR measurements and theory. \\ 

A new ferromagnetic resonance (FMR) method using AMR was developed by
Fang \textit{et al.}~\cite{Fang:2011}. There, an electrical current at
microwave frequencies is used to induce an effective magnetic field in
nanoscale bars of (Ga,Mn)As and (Ga,Mn)(As,P), which is then probed by voltage measurements and analyzed within the framework of non-crystalline AMR~\cite{Fang:2011}. Comparable techniques were employed in order to detect room-temperature spin-orbit torques in the half-Heusler compound NiMnSb~\cite{Ciccarelli:2016_a} and room-temperature spin-transfer torques in a structure consisting of the topological insulator Bi$_2$Se$_3$ and permalloy~\cite{Mellnik:2014}.\\ 
In another context, the AC susceptibility of thin-films of Co, Ni and Nickel alloys was determined by voltage measurements. The expression for the susceptibility (Eq. 6 in Ref. \cite{Booth:2020}) was derived using the non-crystalline AMR~\cite{Booth:2020}.\\ 
And lastly, magnetization reversal was studied by AMR (amongst other means) in Nickel nanowires \cite{Pignard:2000}. However here, the term AMR refers resistance measurements being subject to magnetic field sweeps at different field directions. Jumps in the resistance signal are taken as indication of pinning and unpinning of the magnetic domain walls in the magnetization reversal process. Comparable works can be found in Refs. \cite{Wegrowe:1999, Rheem:2006, Bolte:2005}. A similar study, however with focus on detecting and characterizing the domain wall itself can be found in Ref. \cite{Hayashi:2006}.\\

A central ingredient of the AMR theory is the sd-scattering, which has been mentioned in many positions in this work already. Usually, theoretical predictions of the strength of sd-scattering leading to predictions about the AMR. The opposite is however also possible: A
very small non-crystalline AMR of $0.001\%$ was used to argue that sd-scattering is repressed, the electron carriers and Fermi level reside in the conduction band and the main scattering process is s-s-scattering \cite{PhanNamHai:2012_a, PhanNamHai:2012_b}.\\
While the sd-scattering is governing AMR, its reverse process the $d \rightarrow s$ electron-scattering, is involved in \textit{spin relaxation}. A spin-relaxation theory suitable for nickel- and cobalt-based alloys based on the theory of AMR of Campbell, Fert and Jaoul~\cite{Campbell:1970,Jaoul:1977_a} was developed by Berger and exemplified on permalloy. Parameters of the model were deduced from existing AMR data \cite{Berger:2011}. 

The angular dependence of AMR was used in various occasions: First, in quantifying the current-induced Rashba fields in LAO/STO heterostructures and investigate their dependence on applied magnetic field and on electric field modulation~\cite{Narayanapillai:2014}.
The LAO/STO heterostructures forming a 2DEG, which is more extensively discussed in Sec. 3~\ref{sec_2DEG}.\\
Secondly, the signals of inverse spin Hall effect (ISHE) and of AMR are typically mixed and thus knowledge about AMR is crucial to quantify the spin-Hall angle correctly. In Refs. \cite{Feng:2012_a, Mosendz:2010_a} methods are show how to unwire their signals by symmetry. The ISHE was investigated in permalloy/Pt bilayers~\cite{Feng:2012_a} and in Pt, Au and Mo \cite{Mosendz:2010_a}, respectively.
 \\
AMR can be used to probe the dimensionality of the Fermi surface as was for example done for Ca$_{0.73}$La$_{0.27}$FeAs$_2$ single crystals, where a quasi two-dimensionality of its Fermi surface was found \cite{Xing:2018}.

\subsection{Unconventional Examples and Related Effects}
\label{sec_ApplFurther}

This section attempts to give a short overview on AMR-related research
outside the mainstream, such as the investigation of AMR in non-collinear systems (see sec.~\ref{ssec_ncoll}) where no single spin direction can be defined as in ferromagnets (net magnetization) or in collinear antiferromagnets (N\'eel vector); as well as discussion of similar effects which can partly make use of AMR terminology as such its thermoelectrical counterpart, the Anisotropic Magnetothermopower discussed in sec.~\ref{ssec_AMTP}.

\subsubsection{Frequency-dependent AMR}

This review focuses on AMR in the DC regime. Conductivity is, nevertheless, 
a function of frequency $\sigma(\omega)$ and so is its anisotropy. It
is meaningful to divide the following discussion into low and higher
frequencies. Given the typical scattering rates $1/\tau$ in electrically 
conducting materials, the former means terahertz while the latter
spans the visible range and beyond. In the following paragraph, we discuss AMR in the terahertz regime.\\

The special aspect of the terahertz range is that $\sigma(\omega)$ is
dominated by the intraband contributions which are usually well
approximated by the Drude peak, $\sigma(\omega)\propto (1-i\omega\tau)^{-1}$
where $\tau$ is the transport relaxation time. It is then possible to
split~\cite{Nadvornik:2021_a,Park:2021_a} AMR into
\begin{equation}
	AMR = 
	\frac{\sigma_\perp-\sigma_\parallel}{\sigma_\perp}= \frac{A}{1-i\omega\tau} + B
	\label{eq-04}
\end{equation}
Since the $\omega$-independent term $B$ happens to be a function of the intrinsic AMR and the $\omega$-dependent part of the extrinsic AMR, the ac-AMR offers a possibility for experimentally distinguish these two quantities (see sec. 1\ref{sec_Def} and 3\ref{sec_TM}) \cite{Nadvornik:2021_a}. In these fashion, Co, Ni, Ni$_{50}$Fe$_{50}$~\cite{Nadvornik:2021_a} and permalloy~\cite{Nadvornik:2021_a, Park:2021_a} were investigated (see sec. 1\ref{sec_TM} for the discussion). Please note, that the frequency-dependence was not investigated by means of AC measurements, but instead by means of radiation: The samples were subjected to an incident polarized electrical pulse in THz frequency and the after transmitting through the sample, the outgoing pulse was detected~\cite{Nadvornik:2021_a, Park:2021_a}.\\


Beyond the THz range, interband terms become important, see Eq.~B6 in
Ref.~\cite{Tesarova:2014_a}.  At these higher frequencies
($\omega\tau\gg 1$), the focus turns
to magneto-optical effects which are even in magnetization, such as
the Voigt effect or its analogy in reflection (see Fig.~2 in that
reference for an overview) as counterparts to AMR in the DC-limit.
Spectral measurements than provide information about valence band
structure: iron~\cite{Silber:2019_a}, (Ga,Mn)As~\cite{Tesarova:2014_a}
or Heusler compounds~\cite{Hamrle:2007_a}. Beyond visible and UV region,
core levels can also be probed using x-ray magnetic linear dichroism
(XMLD)~\cite{Valencia:2010_a} but these effects go beyond the scope of
this review. 


\subsubsection{Anisotropic Magnetothermopower}
\label{ssec_AMTP}

The Anisotropic Magnetothermopower (AMTP) is the thermoelectric counterpart of the AMR. Among linear response coefficients
\begin{eqnarray*}
	j  &=& eL_{11}E+ L_{12}\nabla T \\
	j_Q&=& eL_{21}E+ L_{22}\nabla T 
\end{eqnarray*}
it is not only $L_{11}=\sigma/e$ that may depend on magnetization
direction~\cite{Zink:2022_a}. Off-diagonal terms of the $L_{12}$
tensor correspond to the 
Anomalous Nernst Effect  (named in analogy to the anomalous Hall
effect manifested in off-diagonal terms of $L_{11}$), and the AMR (in $L_{11}$)
has the AMTP as its counterpart in $L_{12}$. Magnetoanisotropy
of all these coefficients can be anticipated~\cite{Schlachter:2011}; they
are tensors bound by Onsager relations~\cite{Wegrowe:2013}.

Literature is sparse since measurements and calculations are both challenging.
The measurements of $L_{12}$ are challenging due to possible unwanted
thermoelectric contributions which hardly can be averaged-out
\cite{Ritzinger:2021}. In case of the calculations, the challenge lies
at properly evaluating the derivatives, $L_{12}\sim \int
v_k^2 \delta'(E_k-E_F) $. In the following, we provide a summary of important
literature published on AMTP.\\

Quantitative studies based on phenomenological symmetry-based models analogous to the approach presented in sec. 2\ref{sec_modelpheno} are conducted in Co$_2$MnGa \cite{Ritzinger:2021} and in (Ga,Mn)As \cite{Althammer}. While in Co$_2$MnGa only crystalline and non-crystalline AMTP components up to the second order were confirmed, higher order components have been identified in (Ga,Mn)As. In both cases, AMR and AMTP components were not directly related. A further summary of AMTP studies in (Ga,Mn)As can be found in sec. III - D - 2 of the review by Jungwirth \textit{et al.} \cite{Jungwirth:2014}.

Concerning the $L_{22}$ coefficient, the
AMR was compared by Kimling \textit{et al.} to \textit{anisotropic magnetothermal resistance effect (AMTR)} in polycrystalline Ni nanowires \cite{Kimling:2013} for a range of temperatures. The AMR and AMTR are expressed as ratios and the AMTR is found to be weaker than the AMR due to electron-magnon-scattering.
A two-current model for AMTP in analogy to the work of Campbell, Fert and Jaoul on AMR \cite{Campbell:1970,Jaoul:1977_a} was derived by Heikkil?? \textit{et al}, see Ref. \cite{Heikkila:2010}.
%
%

\subsubsection{Non-collinear systems}
\label{ssec_ncoll}

For a long time, the AMR was only associated with ferromagnets. 
However, the discovery of AMR in collinear antiferromagnets as
described in sec. 3\ref{sec_AFM}, demonstrated that AMR can also
occur in other magnetically ordered materials. Collinear AFM have two
sublattices (typically denoted as \textit{spin up} and \textit{spin down})
which are aligned parallel to each other and allow for the
definition of a single spin axis (the N\'eel vector); practical
differences betweens these two cases, compare the SW1 and SW2
models discussed in Sec. 1(e), are small nevertheless. 
Collinear AFMs alone, however, do not exploit the set of
zero-net-magnetisation systems
to the fullest. In case of non-collinear antiferromagnets, all magnetic moments do point in the same plane, however it is not possible to define a singular spin axis as the Neel vector. An example is magnetic ordering on a kagome lattice in Mn$_3$Sn~\cite{Liu:2017_a} or on a trigonal lattice in CrSe~\cite{CrSe}.
In the latter case, magnetic moments do not lie in the same plane and 
such non-coplanar magnetic order can bring about unexpected consequences.

\begin{figure}
  \hbox{\kern-3cm\includegraphics[scale=0.4,angle=0]{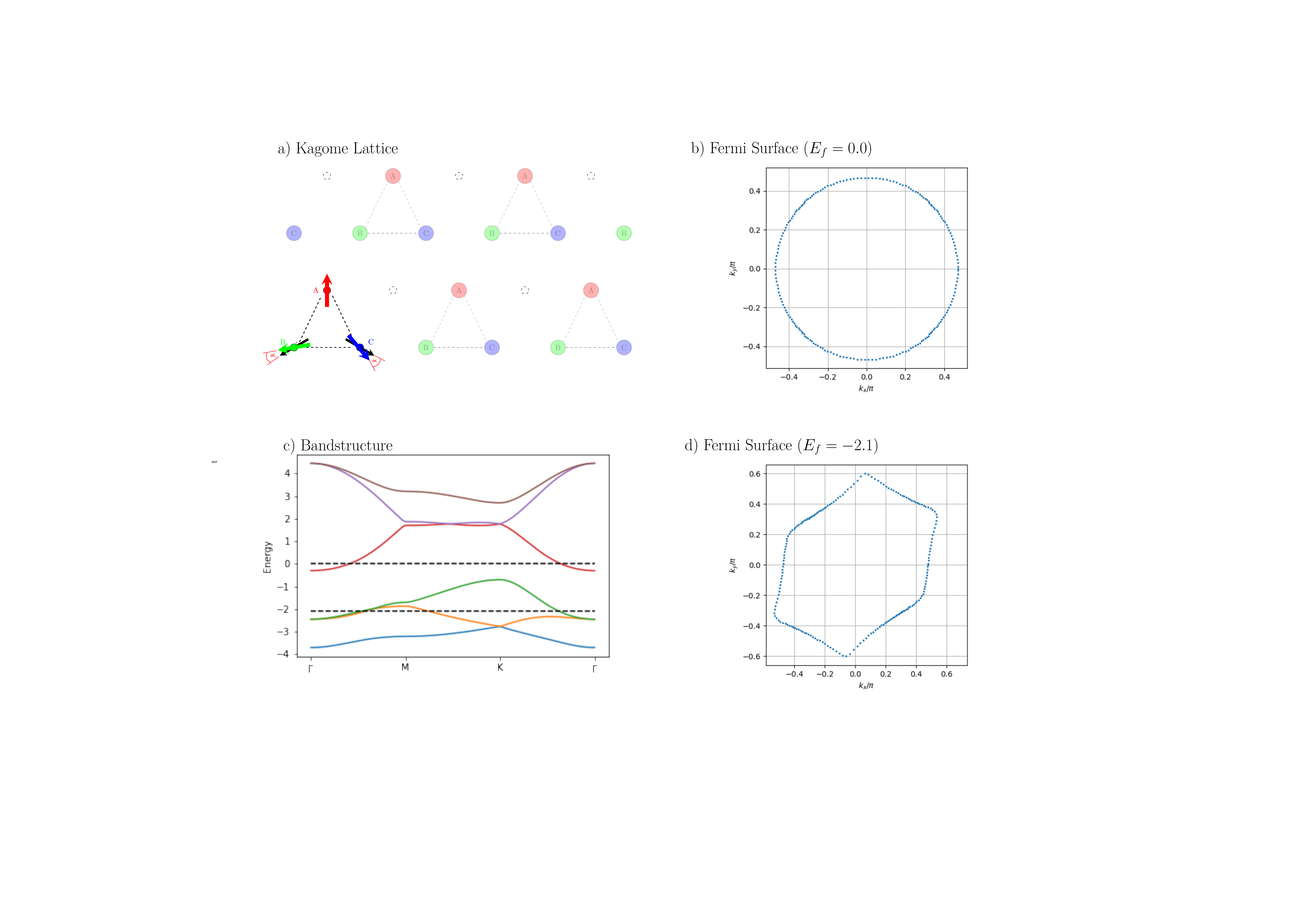}}
  \caption{\lic{YES}{PR+KV}{f16l}
  	Archetypal non-collinear system: kagome lattice with three MSLs.
    (a) Configuration of magnetic moments where (b,d) anisotropy
    occurs depending on $\alpha$. (c) Band structure with Fermi level
    indicated.}
  \label{fig-16}
\end{figure}

Ever since the work of McGuire and Potter~\cite{McG-Potter}, it has
been generally accepted that AMR is an effect relying on the SOI. This
is however only true in collinearly ordered structures. A
non-collinear or non-coplanar order can mimic some properties of the
SOI as, for example, it was shown for AHE on a distorted fcc lattice
endowed with non-coplanar magnetic order~\cite{Shindou:2001_a}.\\
Now, with AMR, non-collinear order is sufficient as we demonstrate in
Fig.~\ref{fig-16}: an s-d model on kagome lattice~\cite{Chen:2014_a}
yields an isotropic band structure as evidenced by Fermi surface (FS)
in panel (b) for a symmetric ($\alpha=0$) configuration of magnetic
moments. We stress that hexagonal warping (when appreciable) does not 
break the isotropy in the sense of $\sigma_{xx}=\sigma_{yy}$ as
discussed in Sec.~2(c). Fig.~\ref{fig-16}(d) shows that for
$\alpha\not=0$ this symmetry is broken and this then leads to intrinsic
AMR even in the absence of SOI. Note that anisotropies in scattering could result in additional extrinsic AMR.\\

Please note that, the definition of AMR could differ between various sources due to the lack of a single spin axis. Some people argue that AMR in non-collinear systems should be due to the simultaneous rotation of all magnetic moments. However this definition is problematic since firstly, the simultaneous rotation of all moments will not change the fermisurface symmetry and thus not allow for any effect and secondly, the application of a magnetic field will not rotate all moments simultaneously. The direction of magnetic moments after application of a magnetic field will be determined by a Stoner-Wohlfarth model SW3,as discussed in sec. 1\ref{sec_stepone}. 

\subsubsection{Exotic Phenomena}
\label{exoticstuff}

In the context of anisotropic magnetotransport, research focuses mostly on the
diffusive regime in bulk systems. In the following, we wish to mention
several phenomena outside this realm 
First, AMR in the ballistic transport regime is discussed, followed by a quick look at tunneling anisotropic magnetoresistance (TAMR) and AMR in topological insulators.

Transport in the diffusive regime is dominanted by scattering, described by the mean-free path of the carrier. When the sample size becomes smaller than the mean-free path we are talking of the ballistic regime: Carriers are only scattered at the boundaries of the sample and can otherwise travel underhindered. Ballistic transport is often times related to one dimensional structures as nanowires. It is possible to have AMR in this regime, which is subsequently either called \textit{ballistic AMR}~\cite{Velev:2005} (BAMR) or \textit{quantized AMR} (QAMR) because of its stepwise character~\cite{Hu:2015_a}. The BAMR is an effect similar to intrinsic AMR since in both cases no external scattering is responsible for the effect. In case of BAMR, the number of bands at the Fermi level and thus the ballistic transport changes with the magnetization direction~\cite{Velev:2005}. It was found that the BAMR is a step function with the magnetization angle~\cite{Velev:2005, Hu:2015_a}. The step-like behavior is only found at low temperatures and for small sample sizes. Increasing the size changes the number of conduction channels and leads to smearing out of the step. An increase of temperature likewise smears out the step~\cite{Hu:2015_a}. In the latter cases, nickel~\cite{Velev:2005} and iron~\cite{Velev:2005, Hu:2015_a} have been investigated.

The dependence of ballistic AMR on in an ideal infinite monoatomic iron wire were compared to influences of domain walls and contacts, both of which can alter the transport properties significantly~\cite{Autes:2008_a}. And lastly, AMR in a Rashba 2DEG was compared between the diffusive and the ballistic regime. The diffusive AMR can be large at low carrier densities which was attributed to the dependence of density of states, while the ballistic AMR shows a nonlinear dependence on the exchange, which was attribtuted to Fermi-surface circle effects~\cite{Kato:2008_a}.\\

Tunneling AMR (TAMR) can be understood as a crossover of AMR, where the anisotropy, thus changing of resistivity with magnetization direction, is important; and the Tunneling Magnetoresistance (TMR)~\cite{Moodera:1995_a, TMR_ref2}, which is based on tunneling as encrypted in its name. Seminal work by Gould \textit{et al.}~\cite{Gould:2004_a} carried out on a structure of a ferromagnetic (Ga,Mn)As layer, a tunneling barrier and a non-magnetic material attributed the anisotropy to the anistropy of the partial DOS~\cite{Gould:2004_a}. 
Since then, TAMR has remained a current topic. Recently, Sch??neberg \textit{et al.}~\cite{JS18}, found that Pb dimers on a ferromagnetic surface show different results depending on the crystalline orientation. For a [001]-oriented dimer, the TAMR reached up to $20\%$, linked to a difference of LDOS depending on the magnetization direction, while TAMR is absent for a [111]-orientation due to only small difference of LDOS depending on magnetiazation~\cite{JS18}.
The TAMR is very much a topic of its own, only related by analogy to the original AMR effect and further discussion is beyond the scope of this review.
\\

Kandala \textit{et al.}~\cite{Kandala:2015_a}
found that in the Cr-doped topological insulator (Bi,Sb)$_2$Te$_3$, which is a FM up to 8~K, exibits a giant AMR of more than $120\%$ in an rotation from out-of-plane to in-plane. This is, because for out-of-plane field, a magnetic gap opens for the surface states
(quantum AHE), but when the field is in-plane, surface states are restored~\cite{Kandala:2015_a}.

\subsection{Industrial Applications}

Argubly the best-known applications of AMR fall into the realm of magnetic memories. Early Magnetoresistive Random Access Memories (MRAM) were based on the effect, yet it is much smaller in magnitude than GMR which eventually prevailed. Modern MRAMs are based on the TMR effect \cite{MRAM_Status}. To date, AMR is still used in applications related to the conventional hard drives~\cite{MRAM_Status} where information is read using a multilayer device~\cite{Bartok:2013_a,Stutzke:2005}
whose resistance changes depending on the magnetic state of the free layer. In the recent decade, numerous attempts of developing novel spintronic applications based on antiferromagnets (see sec. 3\ref{sec_AFM}) were made. In the proposed applications, AMR and its transversal counterpart the planar Hall effect (PHE) were considered as a readout mechanism~\cite{Wadley:2016, Kriegner:2016_a, Marti:2014_a}. To date, AFM spintronic applications did not reach to market readiness. \\

Sensors based on the AMR effect are nowadays still widely used in applications
including detection of absolute position and angle or rotation speed
\cite{Philips_Manual,Honeywell_Handout,Stutzke:2005}. Important is the
usage in the automotive industry, for example in sensing  of crank
shaft position, wheel and transmission speed, throttle valve position
for air intake and many more \cite{Adelerhof:2000}. A list of further
applications can be found in the beginning of Ref.~\cite{Adelerhof:2000}.
Further examples include weak field measurements \cite{Philips_Manual}
such as in a compass \cite{Vcelak:2005}, traffic detection and
measurements of current~\cite{Philips_Manual, Mlejnek:2008}. The measurements 
of current are taking advantage of Ampere's Law where the AMR sensors
detect the magnetic field induced by a current flowing through a wire
\cite{Mlejnek:2008}.
%
AMR sensors offer quite a few advantages, which explains their popularity in industrial applications. They can be produced at low cost \cite{Adelerhof:2000, Honeywell_Handout,Bartok:2013_a}, are quite small \cite{Honeywell_Handout}, achieve a high sensitivity \cite{Philips_Manual, Honeywell_Handout} with resolution well below millimeter or degree-range and are still working if there is a gap between sensor and magnet \cite{Honeywell_Handout} to name only a few. In comparison to Hall sensors they convice with a higher sensitivity \cite{Mlejnek:2008}, lower cost and less sensitivity to mechanical stress \cite{Adelerhof:2000}.\\
%
Requirements for materials used in AMR sensors include large AMR signal (high signal to operating voltage ratio), large $\rho_0$ (noise reduction), low anisotropy, low sensitivity to magnetostriction, long-term stability \cite{Philips_Manual}. Wide temperature ranges are required for operation in e.g. automotive applications as temperature can vary by more than $100^\circ$C. A linear temperature dependence can be compensated electronically \cite{Adelerhof:2000}.
Commonly used materials are mainly basic transition metals discussed in sec.3\ref{sec_TM} and especially permalloy \cite{Philips_Manual,Bartok:2013_a, Schuhl:1994}. The latter has many of the desired properties.\\
%
%
In application-based publications the AMR is identified as non-crystalline AMR \cite{Bartok:2013_a, Philips_Manual, Honeywell_Handout}, treatments of crystalline components are to our knowledge not present. The noise in AMR sensors is typically dominanted by magnetic fluctuations \cite{Stutzke:2005}.
%
%
Please note that similar to AMR sensors, the transverse Planar Hall effect (PHE) can be used to fabricate PHE sensors \cite{Schuhl:1994}.

\section{Conclusion}
\label{ch_Conclusion}

Magnetotransport in solids is a vast and mature field. In this review,
we focus only on a small part of it, namely its anisotropy related to
magnetic order. The anisotropic magnetoresistance (AMR) usually refers to 
this phenomenon albeit occasionally, orbital effects are also included
(and these are not covered in this review). Two characteristic
features of magnetism are helpful to this end: remanence and coercivity.
Unlike ordinary magnetoresistance which just happens to be anisotropic,
the AMR can usually be observed as a spontaneous effect even at zero field;
on the other hand, well above coercive field, magnetoresistance traces
should run in parallel regardless of the experimental configuration
(e.g. magnetic field parallel and perpendicular to current).
Microscopically, the AMR
can either originate from anisotropic scattering or band structure
deformation (related to magnetic order) which is analogous to the 
extrinsic and intrinsic mechanism of the anomalous Hall effect (AHE).
This analogy is not very deep, however, as it can be exemplified with
the intrinsic AMR which is unrelated to Berry curvature of Bloch states.

Phenomenological understanding of the AMR is based on symmetry
analysis (of resistivity tensor) and the basic distinction of
non-crystalline, Eq.~1.2, and crystalline (or mixed) terms allows to
distinguish single crystals from polycrystals where only the former
occurs. Absolute and relative values of the AMR coefficient in Eq.~1.2,
as a material parameter, are useful for polycrystals and single crystals,
respectively. While the latter is usually used, one should be careful:
sputtered films of the same material will exhibit different relative AMR
depending on the strength of scattering on grain boundaries (which is
typically unrelated to magnetism). Strong variations of published AMR values
are therefore to be expected.

While the AHE has attracted considerably more attention than AMR in
fundamental research, situation is quite the opposite in commercial
applications. Contrary to AHE, the AMR has already made it to the
market-ready stage in the niche of various sensors (spintronic
memories, traffic detection and more) and also scientific applications
of AMR (such as a means to determine magnetisation direction in situations
where other methods fail) have become important. More work is needed,
however, to close the gap between real-world applications and the
large body of fundamental research that has been carried out on AMR
over last 165 years.

\section{Acknowledgements}

We wish to express our gratitude to numerous colleagues who
contributed by their comments to writing this review. Support from the
Czech Science Foundation (GA\v CR) via 22-21974S is appreciated.

\pagebreak



\end{document}